\documentclass[11pt]{article}
\usepackage{epsfig,here,hangcaption,rotating,fancyheadings,amssymb,amsmath}
\input epsf
\catcode`\@=11
\def\marginnote#1{}

\def\draftlabel#1{{\@bsphack\if@filesw {\let\thepage\relax
  \xdef\@gtempa{\write\@auxout{\string
    \newlabel{#1}{{\@currentlabel}{\thepage}}}}}\@gtempa
    \if@nobreak \ifvmode\nobreak\fi\fi\fi\@esphack}
     \gdef\@eqnlabel{#1}}
\def\@eqnlabel{}
\def\@vacuum{}
\def\draftmarginnote#1{\marginpar{\raggedright\scriptsize\tt#1}}
\def\draft{\oddsidemargin -.5truein
        \def\@oddfoot{\sl preliminary draft \hfil
        \rm\thepage\hfil\sl\today\quad\militarytime}
        \let\@evenfoot\@oddfoot \overfullrule 3pt
        \let\label=\draftlabel
        \let\marginnote=\draftmarginnote

\def\@eqnnum{(\theequation)\rlap{\kern\marginparsep\tt\@eqnlabel}%
\global\let\@eqnlabel\@vacuum}  }
\def\preprint{\twocolumn\sloppy\flushbottom\parindent 1em
        \leftmargini 2em\leftmarginv .5em\leftmarginvi .5em
        \oddsidemargin -.5in    \evensidemargin -.5in
        \columnsep 15mm \footheight 0pt
        \textwidth 250mmin      \topmargin  -.4in
        \headheight 12pt \topskip .4in
        \textheight 175mm
        \footskip 0pt

\def\@oddhead{\thepage\hfil\addtocounter{page}{1}\thepage}
        \let\@evenhead\@oddhead \def\@oddfoot{} \def\@evenfoot{}
}
\def\titlepage{\@restonecolfalse\if@twocolumn\@restonecoltrue\onecolumn
     \else \newpage \fi \thispagestyle{empty}\c@page\z@
        \def\thefootnote{\fnsymbol{footnote}} }
\def\endtitlepage{\if@restonecol\twocolumn \else  \fi
        \def\thefootnote{\arabic{footnote}}
        \setcounter{footnote}{0}}  
\catcode`@=12
\relax
\newcommand{\newc}{\newcommand}
\newc{\ie}{{\it i.e.}}
\newc{\eg}{{\it e.g.}}

\def\chioi{\tilde{\chi}^0_1}

\def\ohsq{\Omega_{\chi}h^2}

\def\MT2{M_{T2}}

\def\m0{${0}$}

\def\tg{{\tilde g}}
\def\tq{{\tilde q}}
\def\tu{{\tilde u}}

\def\ttop{{\tilde t}}

\def\tchi{{\tilde\chi}}
\def\ttau{{\tilde\tau}}
\def\tl{{\tilde\ell}}

\def\lsp{{\tilde\chi_1^0}}
\def\dofig#1#2{\epsfxsize=#1\centerline{\epsfbox{#2}}}
\def\dofigs#1#2#3{\centerline{\epsfxsize=#1\epsfbox{#2}%
   \hfil\epsfxsize=#1\epsfbox{#3}}}

\begin{document}
\topmargin-1.cm
%
\begin{titlepage}
\vspace*{-64pt}

\begin{flushright}
{YITP-05-76\\
December 2005}\\
\end{flushright}

\vspace{1.8cm}

\begin{center}

\large{{\bf Constraining Dark Matter in the MSSM  at
the LHC}}\\
\vspace*{1.3cm}
\large{M.M. Nojiri}\\
{\it\small YITP, Kyoto University, Kyoto 606-8502, Japan.}\\
\vspace*{0.4cm}
\large{G. Polesello}\\
{\it\small INFN, Sezione di Pavia, Via Bassi 6, I-27100 Pavia,
Italy.}\\
\vspace*{0.4cm}
\large{D.R. Tovey}\\
{\it\small Department of Physics and Astronomy, University of
Sheffield, \\ Hounsfield Road, Sheffield S3 7RH, UK.}\\

\end{center}

\bigskip
\begin{abstract}
In the event that R-Parity conserving supersymmetry (SUSY) is
discovered at the LHC, a key issue which will need to be addressed
will be the consistency of that signal with astrophysical and
non-accelerator constraints on SUSY Dark Matter. This issue is studied
for the SPA benchmark model based on  measurements of
end-points and thresholds in the invariant mass spectra of various
combinations of leptons and jets. These measurements are
used to constrain the soft SUSY breaking parameters at the electroweak 
scale in a general MSSM model. 
Based on these constraints, we assess the 
accuracy with which the Dark Matter relic density can be measured. 
\end{abstract}





\vspace{1cm}

\vspace{-0.5 cm}


\end{titlepage}

\setcounter{footnote}{0}
\setcounter{page}{0}
\newpage

\section{Introduction}
The complementarity of LHC SUSY measurements and direct and indirect
searches for neutralino Dark Matter is an important topic to study
given the increasingly strong astrophysical evidence for cold Dark
Matter in the universe
\cite{Bennett:2003bz,Spergel:2003cb,Peiris:2003ff}. Assuming that
R-Parity conserving SUSY is discovered at the LHC, an interesting
question will arise regarding the compatibility of that signal with
existing relic density constraints (e.g. $0.094<\ohsq<0.129$ at
$2\sigma$ from WMAP data
\cite{Bennett:2003bz,Spergel:2003cb,Peiris:2003ff}), and the
implications it has for terrestrial Dark Matter searches.

In a previous paper \cite{Polesello:2004qy}
we addressed these issues within the context of the
minimal supergravity (mSUGRA) class of SUSY models incorporating
gravity-mediated SUSY breaking \cite{msug}. There is a great advantage in 
studying models such as the mSUGRA ones, where a definite pattern 
of SUSY breaking is imposed at the unification scale. These models 
are in fact described with only a small number of independent parameters.
It is therefore possible, through a limited number of measurements
to fully constrain the model, and extract very precise predictions
for physical quantities related to Dark Matter.
However, the mSUGRA model is strongly constrained by the WMAP 
data,  and we do not know whether  from the LHC measurements an unambiguous
determination of the high scale behaviour of the SUSY theory 
will be possible. 
It is therefore interesting to study a generic model 
in which the weak-scale SUSY breaking parameters are 
independent, such as the Minimal Supersymmetric Standard Model
(MSSM), and verify to what level 
the LHC measurements can constrain the neutralino relic density 
and the cross-sections for direct Dark Matter detection.\par
Some recent papers \cite{Battaglia:2005ie, Birkedal:2005jq},
in the framework of International Linear Collider (ILC)  studies, based on 
generic and sometimes restrictive assumptions on the measurement
potential of the LHC, conclude for a very limited potential
of the LHC experiments in constraining the neutralino relic 
density in a generic MSSM. The aim of this paper is to investigate in detail, 
by carefully taking into account published studies, 
what are the effective limitations of the LHC measurements
in this field for a specific representative model for which 
sufficient information is available. 
 
We choose to study one particular point of the MSSM 
parameter space, which was adopted as a benchmark point 
by the SPA group \cite{Aguilar-Saavedra:2005pw}. This model is defined in terms 
of the parameters of the mSUGRA model 
($m_0=70$ GeV, $m_{1/2}=250$ GeV, $A_0=-300$ GeV,
$\tan{\beta}=10$, $\mu>0$). This is a modification 
of the point SPS1a,  essentially achieved 
by lowering $m_0$ from 100 to 70~GeV,  originally defined in
Ref.~\cite{Allanach:2002nj}, to take into account
the WMAP results. It is also very near to Point B$^\prime$
of \cite{battaglia03}.  This model lies in the `bulk' region of
the $m_0-m_{1/2}$ mSUGRA plane where the relic density is reduced to a
small value by $\chioi$ annihilation to leptons or neutrinos via
$t$-channel slepton, stau and sneutrino exchange. 
The values of the MSSM soft parameters 
and of the sparticle  masses were computed with the program ISASUSY 7.71
\cite{isa},
taking the tree-level values for the sparticle masses, and
are given in Table~\ref{tab:masses}.
The Lightest Supersymmetric Particle (LSP) relic density was 
computed based on these inputs 
using the program micrOMEGAs 1.3.6 \cite{Belanger:2001fz}, 
and the resulting value is 
\mbox{$\Omega_\chi h^2$=0.108}, well within the WMAP range. 
The main annihilation processes contributing to the calculation of 
the relic density are given  in Table~\ref{tab:anni}. 
As discussed above we take this point as a benchmark, but we analyze
it without assuming a predefined relationship among the 
SUSY breaking parameters.

In Section 2 we describe the LHC measurements which can be used
to constrain SUSY at the considered model  point.
In Section 3 we develop a strategy to constrain the MSSM parameters
relevant to the calculation of the neutralino relic density.
In Section 4 we discuss the
use of these constraints to calculate the $\chioi$
relic density.
Finally in Section 5 we review the
results and discuss the general applicability of the technique.
\begin{table}[htb]
\begin{center}
\vskip 0.2cm
\begin{tabular}{|c|c|c|c|}
\hline
Sparticle & mass (GeV) & Sparticle & mass (GeV) \\
\hline
\hline
$\lsp$ & 97.2 & $\tchi^0_2$ & 180.1 \\
$\tchi^0_3$ & 398.4 & $\tchi^0_4$ & 413.8 \\
$\tl_L$ & 189.4 & $\tl_R$ & 124.1 \\
$\ttau_1$ & 107.7 & $\ttau_2$ & 194.2 \\
$\ttop_1$ & 347.3 & $\ttop_2$ & 562.3 \\
$\tu_L$ & 533.3 & $\tg$ & 607.0 \\
$h$ & 116.8 & $A$ & 424.6 \\
\hline
\end{tabular}
\caption{\label{tab:masses}  {\it Masses of the sparticles
in the considered model as calculated at tree level
with ISAJET 7.71 \cite{isa}
}}
\end{center}
\end{table}

\begin{table}[htb]
\begin{center}
\vskip 0.2cm
\begin{tabular}{|l|r|}
\hline
Process  & Fraction \\ 
\hline
\hline
$ \lsp \lsp \rightarrow \ell^+ \ell^-$  & 40\% \\ 
$ \lsp \lsp \rightarrow \tau^+ \tau^-$  & 28\% \\ 
$ \lsp \lsp \rightarrow \nu \bar{\nu}$   & 3\%  \\  
$ \lsp \ttau_1 \rightarrow Z \tau$      & 4\%  \\ 
$ \lsp \ttau_1 \rightarrow A \tau$      & 18\% \\  
$ \ttau_1 \ttau_1 \rightarrow \tau\tau$ &  2\% \\  
\hline
\end{tabular}
\caption{\label{tab:anni} {\it Fractional contribution of different annihilation
processes to the LSP relic density in the considered model.
The relic density was calculated with the program 
micrOMEGAs 1.3.6 \cite{Belanger:2001fz}.
}}
\end{center}
\end{table}

\section{The measurement of SUSY parameters at the LHC}
\label{Sec:meas}

The development of techniques for measuring parameters characterizing
SUSY models has been the subject of much investigation in the last few
years, as documented in Ref.~\cite{Bachacou:2000zb,atltdr,
Allanach:2000kt}, and is still a very active field of investigation.

The basic issue is that the presence of two invisible particles in the
final state renders the direct measurement of sparticle masses through
the detection of invariant mass peaks impossible. Alternative
techniques have therefore been developed, based on the exclusive
identification of long cascades of two body-decays. It was
demonstrated \cite{Bachacou:2000zb,atltdr} that if a sufficiently long chain
can be identified (at least three successive two-body decays), the
thresholds and end-points of the various possible invariant mass
combinations among the identified products can be used to achieve a
model-independent measurement of the masses of the involved
sparticles. Once the masses of the lighter sparticles are obtained
with this procedure, in particular the mass of the LSP,
additional sparticle masses can be
measured through the identification of other shorter decay
chains. This program has been carried out recently for the SPS1a 
model point \cite{LHCLC}, assuming the performance
of the ATLAS detector. This study 
results in a number of measurements of observables
which are related to the masses of the sparticles by known algebraic
relations. 
For the work presented here, we assume for the corresponding
observables in the SPA model the same errors as the ones
obtained in the full analysis for Point SPS1a. 
The position of the kinematic edges is quite similar 
in the two points, due to the fact that only the slepton
spectrum is modified by the $30\%$ change in $m_0$, and 
the total production cross-section is somewhat higher
in the new point, which should result is somewhat 
smaller statistical errors. 
We further checked that
the mass spectrum for the SPA model does not present 
degeneracies  in the sparticle masses which would 
severely reduce the transverse momenta of the visible decay
products and thence the experimental selection efficiency.
The considered measurements are give in 
Table~\ref{tab:summes}. The meaning of the different
observables, and their expression in terms of sparticle masses is
given in Ref.~\cite{Allanach:2000kt}. The scale error is derived 
from the assumption of a control on the lepton energy scale 
at the level of 0.1\%, and of the jet energy scale at the level
of 1\% \cite{atltdr}. 
Following \cite{Gjelsten:2004ki}, the 1\% jet scale uncertainty
reflects in a $\sim 0.5\%$ uncertainty on the position of the edges 
involving jets and leptons.

\begin{table}[htb]
\begin{center}
\vskip 0.2cm
\begin{tabular}{|l|c|c|c|c|}
\hline
& & \multicolumn{3}{c|}{Errors}\\
Variable & Value (GeV) & Stat+Sys (GeV) & Scale (GeV) & Total \\
\hline
\hline
$m_{\ell\ell}^{max}$             &    81.2 &   0.03 &    0.08 &  0.09 \\
$m_{\ell\ell q}^{max}$           &   425.3 &    1.4 &    2.1 &    2.5 \\
$m_{\ell q}^{low}$               &   266.9 &    0.9 &    1.3 &    1.6 \\
$m_{\ell q}^{high}$              &   365.9 &    1.0 &    1.8 &    2.1 \\
$m_{\ell\ell q}^{min}$           &   207.0 &    1.6 &    1.0 &    1.9 \\
$m(\ell_L)-m(\lsp)$              &    92.3 &    1.6 &    0.1 &    1.6 \\
$m_{\ell\ell}^{max}(\tchi^0_4)$  &   315.8 &    2.3 &    0.3 &    2.3 \\
$m_{\tau\tau}^{max}$             &    62.2 &    5.0 &    0.3 &    5.0 \\
\hline
\end{tabular}
\caption{\label{tab:summes} {\it Summary table of the SUSY
measurements which can be performed at the LHC with the ATLAS
detector \cite{LHCLC} used in this analysis. 
The central values are calculated with ISASUSY 7.71, 
using the tree-level values for the sparticle masses.
The statistical errors are given for the integrated
luminosity of 300~$fb^{-1}$. The uncertainty in the energy scale is
taken to result in an error of 0.5\% for measurements including jets,
and of 0.1\% for purely leptonic mesurements.}} 
\end{center}
\end{table}

Some additional measurements are considered in this analysis:
\begin{itemize}
\item
The mass of the
light Higgs boson from the decay \mbox{$h\rightarrow\gamma\gamma$}
which for 100~fb$^{-1}$, and for $m_h=114$~GeV, has a statistical
uncertainty of $\sim0.5$~GeV \cite{hohlfeld}.
\item
The ratio of branching ratios:
$$
BR(\tilde\chi^0_2\rightarrow\tilde{\ell_R}\ell)/
BR(\tilde\chi^0_2\rightarrow\ttau_1\tau).
$$  
No detailed experimental study is available on this measurement.
From a particle-level evaluation of the number of events involved, the
error on this quantity will be dominated by the systematic uncertainty
in the experimental efficiency for the $\ttau_1$ channel, in
particular near threshold where the contribution to the visible ditau
mass distribution is significant.  We assume here a conservative error
of 10\%.
\item
The constraints from searches for heavy Higgs bosons of 
MSSM, which will be discussed in detail in the following.
\end{itemize}

Two different types of uncertainties are quoted in Table
\ref{tab:summes}: the combined statistical and systematic
uncertainties estimated for each measurement, and general errors on
the scales of lepton and hadron energy measurement, which affect all
the measured quantities in the same way. Since in many cases the scale
uncertainties are dominant it is necessary to take into account the
correlations between the different measurements when extracting the
constraints on the model parameters. 
In order to take into account the correlations we 
use a Monte Carlo
technique relying on the generation of simulated experiments sampling
the probability density functions of the measured observables. In
frequentist statistics, confidence bands describe the probability that
an experiment in a set of identical experiments yields a given value
for the measured quantities.

\par We proceed in the following way:
\begin{enumerate}
\item
An `experiment' is defined as a set of measurements, each of which
is generated by picking a value from a gaussian distribution with mean
given by the central value given in Table~\ref{tab:summes}. The
correlation due to energy scale is taken into account by using a
second gaussian distribution for the energy scale, and using the same
random number for all the measurements sharing the same scale.
\item
For each experiment, we extract the constraints on the MSSM 
model as we will describe in the following.
\end{enumerate}

We obtain as a result of this calculation a set of MSSM models, each
of which is the ``best'' estimate for a given Monte Carlo experiment of the model generating the observed measurement pattern. For each of
these models the properties of the LSP Dark Matter candidate and other
SUSY particles involved in Dark Matter annihilation can then
be calculated, with the spread of obtained results being interpreted as the
level of precision with which these properties can be measured by the
LHC. As we are working in a general MSSM, some of the parameters will
only be loosely constrained, if at all, by the measurements.
The spread obtained from the variation of the unconstrained parameters
needs to be studied in detail to assess the final prediction precision on
DM characteristic.

\section{Extraction of MSSM parameters}
\label{sec:mssm}
\begin{figure}[htb]
\begin{center}
\dofigs{0.5\textwidth}{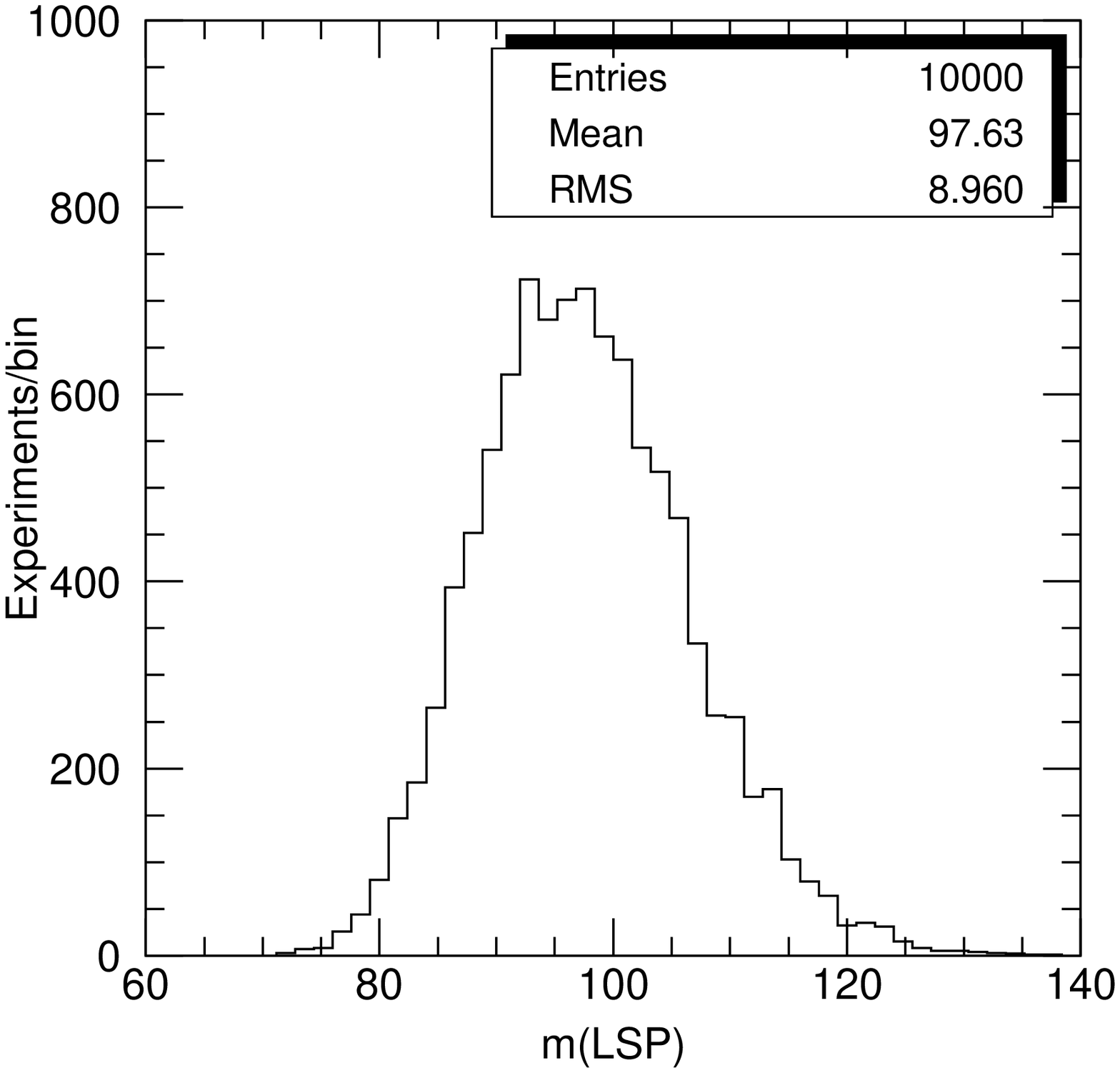}{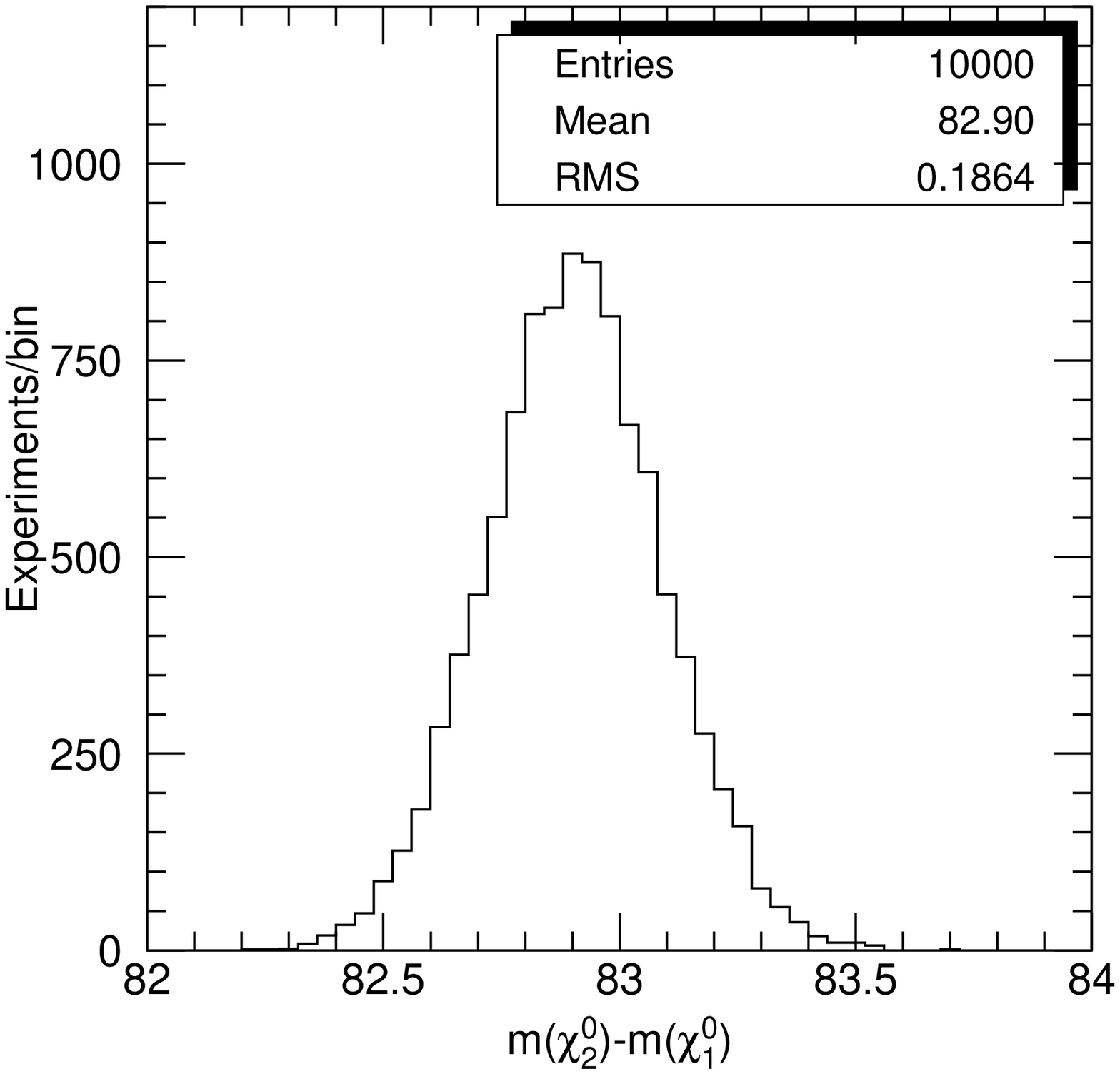}
\caption{\label{fig:mchi01} {\it
Distribution of the measured value of $m(\lsp)$ (left) 
and of the difference $m(\tchi^0_2)-m(\lsp)$ (right)
for a set of Monte Carlo
experiments, each corresponding to an integrated statistics of
300~fb$^{-1}$.}}
\end{center}
\end{figure}

In order to extract the parameters of the MSSM we proceed in 
a stepwise fashion. We first extract the measurement of the
sparticle masses from the measured edges. 
We do not address here the issue of possible ambiguities which 
can arise on the identity of the particles involved in the
decay chains as discussed in \cite{Lester:2005je} and \cite{Gjelsten:2005sv}.
The procedure yields an error 
of $\sim$9~GeV for the masses of the sparticles relevant to this study.
We show in the left side of 
Figure~\ref{fig:mchi01} the distribution of the measured 
$\lsp$ masses for a set of Monte Carlo experiments. 
The error on the mass is basically independent
on the mass of the sparticle  because the errors 
on masses are strongly correlated. This is because of the algebraical 
structure of the formulas in \cite{Allanach:2000kt},
\cite{Gjelsten:2004ki}, which causes
the absolute scale to be much more loosely constrained than the 
difference between masses. 
In particular, compared to the absolute error of $\sim9$~GeV 
quoted above, the error on the difference $m(\tl_R)-m(\tchi^0_1)$ 
is $\sim200$~MeV , as shown on the right side
of Figure~\ref{fig:mchi01}, thanks to the excellent precision of the measurement
of the $m_{\ell\ell}^{max}$ edge position. 
The calculated precision on
$m(\ttau_1)-m(\tchi^0_1)$ is $\sim2.5$~GeV, scaling approximately
linearly with the assumed error on $m_{\tau\tau}^{max}$.\par
After extracting the mass measurements, the next step consists of 
constraining the neutralino sector, and calculating the
components of  the neutralino mixing matrix, which is a necessary
ingredient in all calculations on the neutralino annihilation rate.
Once this is fixed, we go on to constrain the slepton sector.
The next step is understanding the constraints in the Higgs sector
in order to evaluate possible contributions to the annihilation 
rate of Higgs exchange diagrams. 
By putting all of the information together we are finally able 
to give an estimate on the predictive  power of the LHC data
for a specific scenario in which the neutralino annihilation 
is dominated by processes involving sleptons.

\subsection{Constraints on the neutralino mass matrix}
\begin{figure}[htb]
\begin{center}
\dofigs{0.5\textwidth}{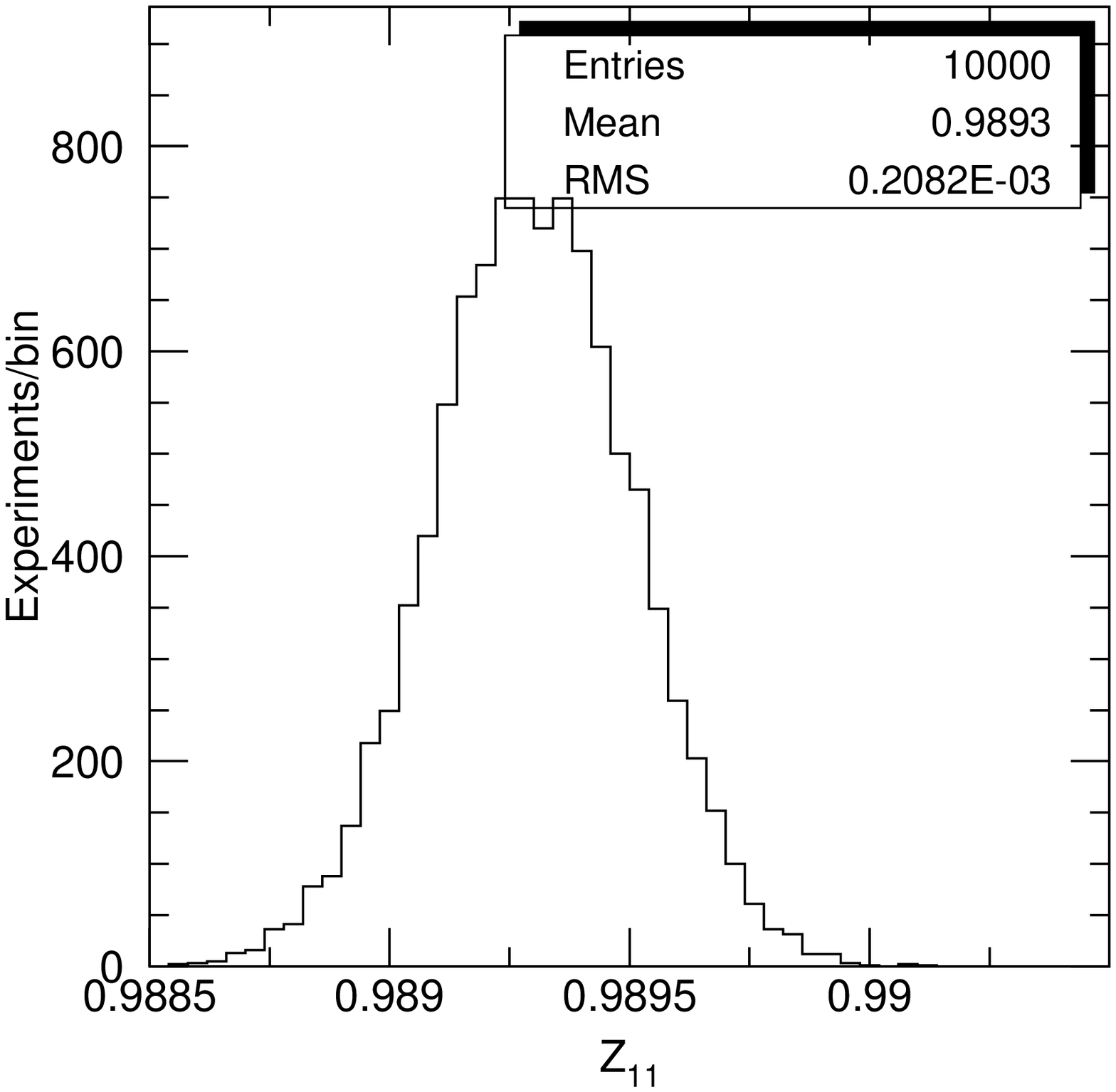}{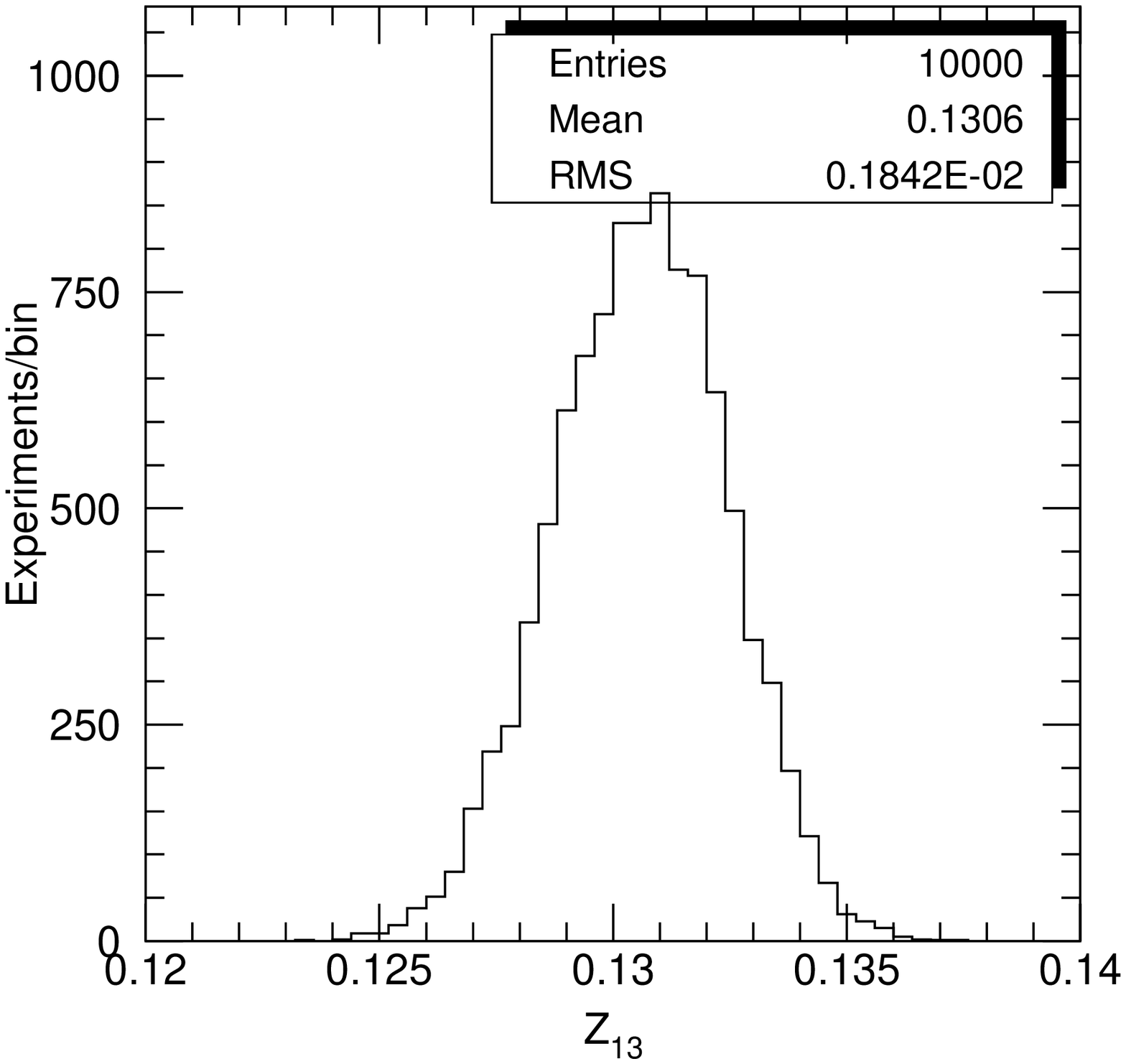}
\caption{\label{fig:z11z13} {\it Distributions of 
the measured values of  $Z_{11}$, 
the bino component of $\tchi^0_1$ (left)
and of $Z_{13}$, the Higgsino$_1$ component of $\tchi^0_1$ (right)
assuming a fixed value $\tan\beta=10$ for a set of Monte Carlo 
experiments, each corresponding to an integrated statistics of 
300~fb$^{-1}$.}}
\end{center}
\end{figure}

Based on the expected LHC measurements, the masses of three 
neutralinos are measured: $\tchi^0_1$, $\tchi^0_2$, $\tchi^0_4$.
In the MSSM, assuming
a $CP-$conserving scenario, the mass eigenstates
$\chi_i^0$ ($i$=1,2,3,4) result from the mixing of 
the SUSY partners of the neutral gauge and Higgs  bosons:
$$(\tilde{B},\tilde{W}^3,\tilde{H}^0_1,\tilde{H}^0_2)$$
through a mixing matrix defined as:
\begin{eqnarray}
{\cal M}=\left(\begin{array}{cccc}
  M_1       &      0          &  -m_Z \cos\beta s_W  & m_Z \sin\beta s_W \\[2mm]
   0        &     M_2         &   m_Z \cos\beta c_W  & -m_Z \sin\beta c_W\\[2mm]
-m_Z \cos\beta s_W & m_Z \cos\beta c_W &       0       &     -\mu        \\[2mm]
 m_Z \sin\beta s_W &-m_Z \sin\beta c_W &     -\mu      &       0
                  \end{array}\right)\
\label{eq:massmatrix}
\end{eqnarray}
The above matrix 
is built out of four MSSM parameters: $M_1$ and $M_2$ are respectively 
the U(1) and SU(2)
gaugino masses, $\mu$ is the Higgsino mass parameter, $\tan\beta=v_2/v_1$
is the ratio of the vacuum expectation values of the two Higgs
doublets of the model. The additional parameters, $s_W$ and $c_W$,  
respectively the sine and cosine of the electroweak
mixing angle $\theta_W$, and $m_Z$, the mass of the $Z$ boson are precisely 
known from measurements in the Standard Model.
Given that we only have three input parameters in the three masses, 
we miss one parameter to fully solve the neutralino matrix.
We take this parameter to be $\tan\beta$, as the other parameters
relate directly to the neutralino masses. 
We show in Figure~\ref{fig:z11z13} the distribution of the
reconstructed values of $Z_{11}$  and $Z_{13}$, for $\tan\beta=10$,
where the composition
of the $\tchi^0_1$ is written as:
$$
\tchi^0_1= Z_{11} \tilde{B} + Z_{12} \tilde{W}^3 + Z_{13} \tilde{H}^0_1
+ Z_{14} \tilde{H}^0_2
$$
\begin{figure}
\begin{center}
\dofig{\textwidth}{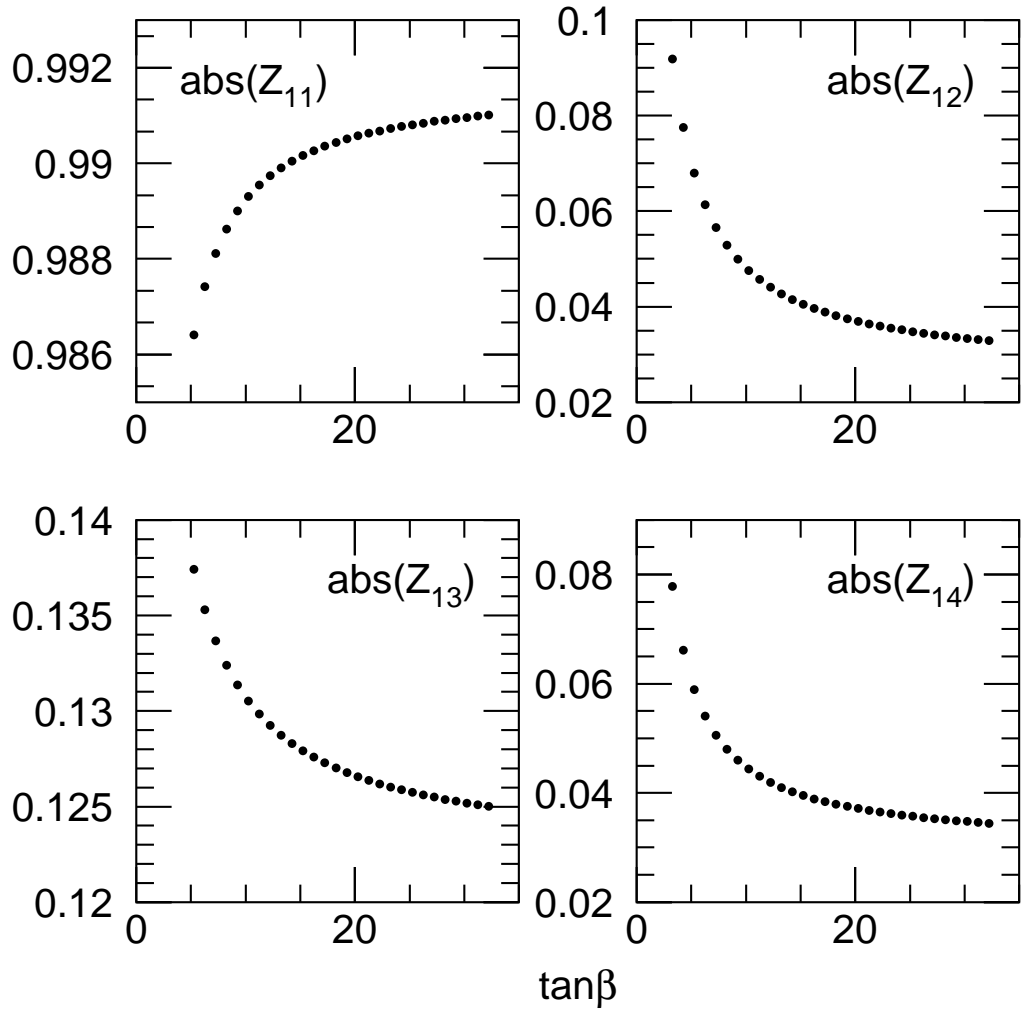}
\caption{\label{fig:z4dep} {\it 
Average values of the  four components
of $\tchi^0_1$ 
as a function of the assumed value of $\tan\beta$.
The averages are  performed  on sets of Monte Carlo
experiments, each corresponding to an integrated statistics of
300~fb$^{-1}$.}}
\end{center}
\end{figure}
The spread from the experimental error is 0.02\% for the bino component,
and 1.5\%  for the other  components. The dependence of 
the four components of the LSP on 
the assumed value of $\tan\beta$ is shown in Figure~\ref{fig:z4dep}, 
and for the assumed range 3-30 is 0.8\% for $Z_{11}$ and $\sim$15\% 
for $Z_{13}$. A much larger variation with $\tan\beta$ is observed for the 
smaller components $Z_{12}$ and $Z_{14}$.
The sensitivity is higher for low values of $\tan\beta$. 
A first consideration from this measurement is that already at
this stage the data tell us that the neutralino  is an almost
pure bino, with only a small Higgsino component.
Annihilation channels involving the exchange
of sfermions in the $t$-channel will dominate, 
unless the masses of the heavy Higgses are such that they resonantly
enhance the relevant annihilation processes.\\
In the calculation above $M_1$ and $M_2$ are assumed as independent, 
since we are working on a general MSSM model. It is however customary
in MSSM studies to assume that the ratio of the gaugino masses 
is equal to the ratio of the respective coupling 
constants, giving the relation $M_2=5/3\tan^2\theta_WM_1$. 
Under this assumption we can invert the gaugino mixing matrix
using as constraints the mass differences $m(\tchi^0_2)-m(\lsp)$
and $m(\tchi^0_4)-m(\lsp)$ instead of the measured masses,
and extract the values of $M_2$ and $\mu$, always for fixed $\tan\beta$.
From these one can in turn calculate the values of the gaugino 
masses. In this case the absolute scale is fixed in a much more
precise way, and the resolution on $m(\lsp)$ is $\sim 250$~MeV.
We will keep as a baseline for our study the completely unconstrained
model, but we will also show the results when the gaugino
mass unification is assumed. 
\subsection{Constraints on the slepton sector}
Once the neutralino mass is extracted, the slepton sector can 
be considered. No mixing is assumed here in the selectron and
smuon case. It was shown in \cite{Goto:2004cp} that 
the difference in mass of the electron and muon should result
in different mixings for the selectron and the smuon, resulting
for the considered model point
in a small mass difference and in a $\sim$8\% difference 
between $BR(\tilde\chi^0_2\rightarrow\tilde{e}_R e)$ and
$BR(\tilde\chi^0_2\rightarrow\tilde{\mu}_R\mu)$.
We assume for this analysis that no mixing 
takes place in the selectron and smuon sector.\par
Due to the hypercharge difference, 
the LSP pair annihilation cross section through 
$t$-channel exchange of $\tl_R$   is 16 times bigger than 
that for $\tl_L$  exchange when $m(\tl_R)_R=m(\tl_L)$  and 
the LSP is dominantly bino. 
Therefore if the chirality of the slepton 
observed in the $\tchi^0_2$ decay chain can not 
be determined, this could cause a very large uncertainty 
in the relic density calculation.
The issue is however discussed in \cite{Goto:2004cp},
where it is shown that the asymmetry distribution proposed in 
\cite{Barr:2004ze} is sensitive to the chirality structure of the slepton.
Those studies were performed for the SPS1a point, so their results
can be safely extended to the model addressed in the present study,
and we can assume for the present study that the chirality of 
the lighter slepton can be determined.\par
\begin{figure}[htb]
\begin{center}
\dofigs{0.5\textwidth}{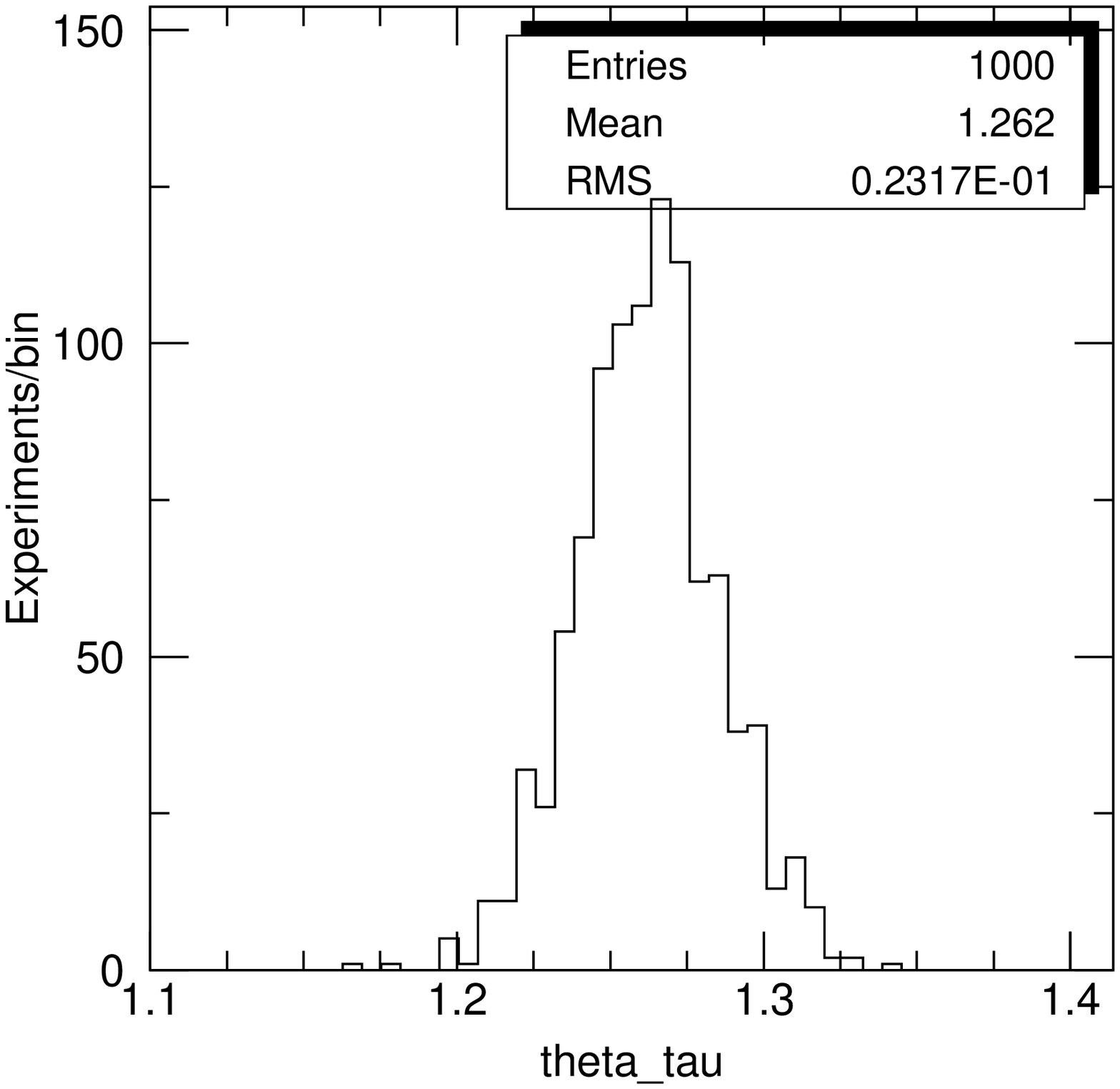}{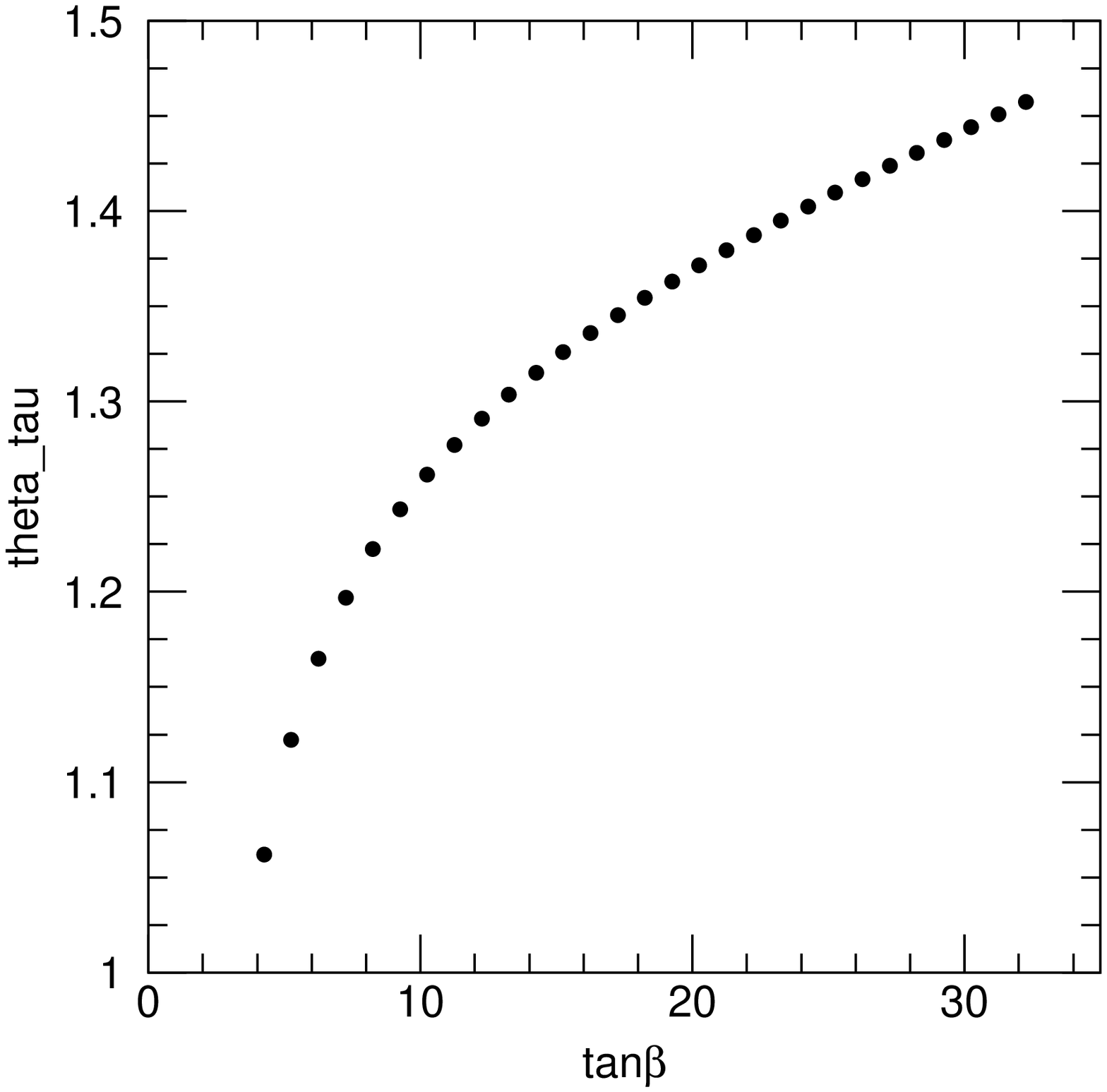}
\caption{\label{fig:thetatautb} {\it
Left: distribution of the measured value of $\theta_\tau$,
the $\ttau$ mixing angle 
assuming a fixed value $\tan\beta=10$ for a set of Monte Carlo
experiments, each corresponding to an integrated statistics of
300~fb$^{-1}$.
Right: average values of $\theta_\tau$, 
as a function of the assumed value of $\tan\beta$.
The averages are  performed  on  sets of Monte Carlo
experiments, each corresponding to an integrated statistics of
300~fb$^{-1}$.}}
\end{center}
\end{figure}

In the  stau sector, the ratio 
\mbox{$ BR(\tilde\chi^0_2\rightarrow\tl_R\ell)/
BR(\tilde\chi^0_2\rightarrow\ttau_1\tau) $} 
is a function of the neutralino mixing matrix, of 
$m(\ttau_1)$, $m(\tchi^0_2)$,  $\tan\beta$ and  $\theta_\tau$, 
the mixing angle between $\ttau_R$ and $\ttau_L$. 
By assuming a given value for $\tan\beta$ we can therefore 
from this measurement extract the value of $\theta_\tau$.
The distribution of the measured $\theta_\tau$, as 
well as the  dependence on the input 
$\tan\beta$ value are are shown in Figure~\ref{fig:thetatautb}.
The experimental uncertainty is $\sim$2\%, whereas 
$\theta_\tau$ varies by $\sim 35\%$ over the considered $\tan\beta$ range.

The only missing parameter in order to be able 
to calculate the neutralino annihilation processes involving sleptons 
is the mass of the $\ttau_2$ slepton. 
Natural bounds can be imposed to this mass.
First of all we must have $m(\ttau_2)>m(\tchi^0_2)-m(\tau)$, 
otherwise the $\ttau_2$ contribution would likely be observed 
in the $\tau\tau$ edge from the $\tchi^0_2$ decay.
Second, from $m(\ttau_1)$, $\theta_\tau$, $\tan\beta$ and
 $m(\ttau_2)$ the value for the trilinear coupling 
$A_\tau$ can be calculated. Large values of the $A_\tau$ parameter
could  induce charge breaking minima (CCB) due to the vacuum 
expectation values of the charged $\tau$ scalars. Typical 
constraints on $A_\tau$ from these considerations \cite{Casas:1995pd}
would give an upper limit on $A_\tau$  of $\sim$500~GeV, 
very near to the actual $A_\tau$ value for the considered model. 
The conditions on CCB minima resulting in constraints on the MSSM 
are in general neither necessary nor sufficient to give
an acceptable vacuum structure to the theory, and it
has been suggested \cite{Kusenko:1996jn} that the constraint from 
\cite{Casas:1995pd} be relaxed through an empirical multiplicative 
factor. Following these considerations we conservatively fix 
here a limit of 5~TeV to  $A_\tau$, which, in the
considered model, for $\tan\beta=10$ corresponds to a limit
$m(\ttau_2)<250$~GeV.  
\subsection{Constraints from the Higgs sector}
\begin{figure}[htb]
\begin{center}
\dofig{0.7\textwidth}{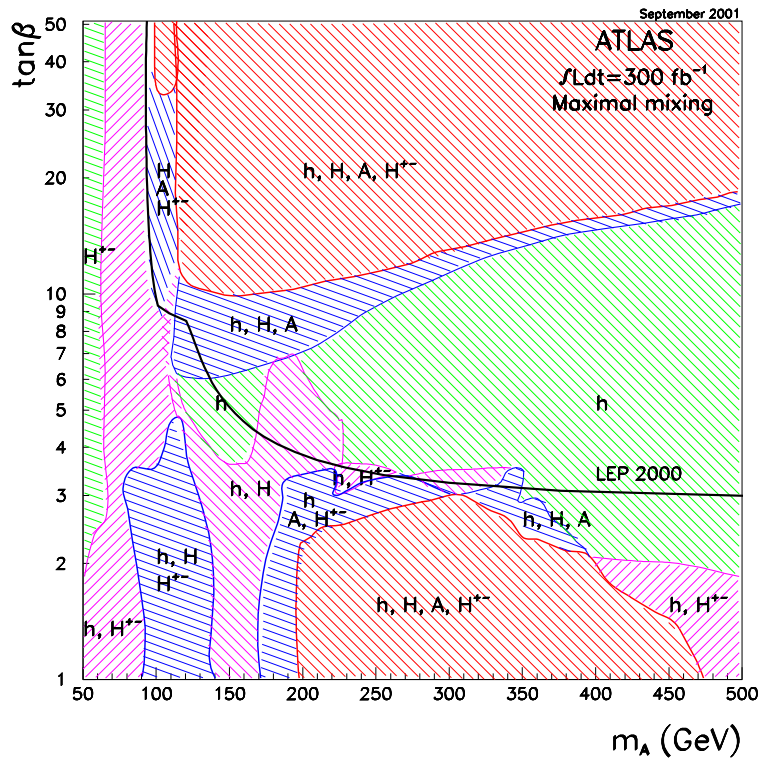}
\caption{\label{fig:mssmatlas} {\it
Reach of the ATLAS experiment in the \mbox{$m(A)-\tan\beta$}
plane for an integrated luminosity of 300~fb$^{-1}$. For each region
in the plane, the detectable Higgs bosons are marked.
}}
\end{center}
\end{figure}

As discussed in the previous sections, the measurements considered up to
now do not allow us to constrain the $\tan\beta$ parameter.
The main constraints come from the Higgs sector.
In particular, the measurement of the mass and of the production
rate of one of the heavy Higgs would define a confidence region  in the 
canonical $m(A)-\tan\beta$ plane.
The region where heavy Higgses can be discovered through their decays to 
Standard Model particles are shown in Figure~\ref{fig:mssmatlas}, 
from \cite{atltdr}.

The plot was obtained on specific assumptions on the SUSY spectrum,
i.e. a SUSY mass scale $M_{SUSY}$ of 1~TeV, and maximal mixing, i.e. 
$\mu\ll M_{SUSY}$ and the trilinear couplings $A=\sqrt{6}M_{SUSY}$.
These assumptions mostly affect the considerations
related to the light Higgs, 
and only lightly affect the discovery region for H/A and $H^+$.
Given that the mixing in the chosen model is less than maximal
($A_t=-450$~GeV), the limit on $\tan\beta$, $\tan\beta>3$ 
shown in Figure~\ref{fig:mssmatlas} can be used as lower limit
of the $\tan\beta$ range considered in this analysis.
The considered model, corresponding to $m_A=424$~GeV and 
$\tan\beta=10$ lies outside of the region where a heavy Higgs 
can be discovered, we will limit our analysis to the
complement of this region in the $m(A)-\tan\beta$ plane.\par
A stronger constraint can be obtained by the fact that
the light Higgs $h$ will be discovered at the LHC and its
mass measured. From this measurement a confidence band in 
the $m(A)-\tan\beta$ plane can be defined.\par
Finally, since the SUSY spectrum is largely known, we can investigate
whether the heavy Higgses can be detected either a) in the cascade
decay of a sparticle, or b) through their decay into a sparticle,
or a bound can be extracted from the non-observation.\par

\begin{figure}[htb]
\begin{center}
\dofig{0.6\textwidth}{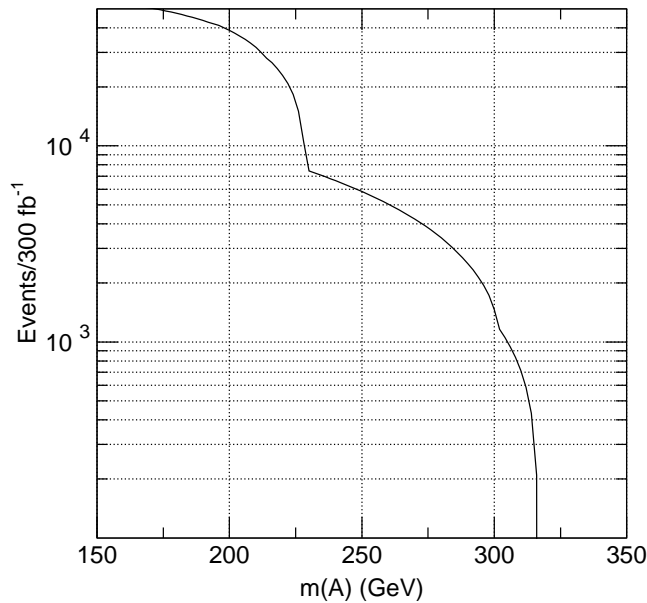}
\caption{\label{fig:plothb} {\it
Number of expected events containing a \mbox{$H/A\rightarrow b\bar{b}$}
decay in the cascade of squarks or gluinos, as a function 
of $m(A)$. The assumed integrated luminosity is 300~fb$^{-1}$
}}
\end{center}
\end{figure}

For the cascade decays, no heavy Higgs appears in the decay chains,
as all the decays of neutralinos/charginos into heavy Higgses
are kinematically closed. We can however investigate up to which 
Higgs masses these decays would be open.
The best candidates would be the decays: 
$\tchi_{4(3)}^0\rightarrow\tchi^0_{1(2)} A/H$,  
$\tchi^\pm_2\rightarrow\tchi^\pm_1 A/H$ with subsequent decays 
of the A/H into $b\bar{b}$. 
These decays have been studied in \cite{Datta:2003iz}
where it is shown that in favourable conditions a peak in  $b\bar{b}$
distribution can be observed. The kinematic limits in the considered model
are:  
\begin{itemize}
\item
$m(A/H)\le 315$~GeV \, \, for 
\mbox{$\tchi^0_{4(3)}\rightarrow\tchi^0_{1} A/H$}
\item
$m(A/H)\le 230$~GeV \, \, for 
\mbox{$\tchi^0_{4(3)}\rightarrow\tchi^0_{2} A/H$} and
\mbox{$\tchi^\pm_2\rightarrow\tchi^\pm_1 A/H$}.
\end{itemize}
We evaluated the number of events in which a $H/A\rightarrow b\bar{b}$
decay is produced in the cascade decay of 
squarks and gluinos by generating with HERWIG \cite{HERWIG}
a sample of events 
for the chosen model, and computing the fraction of events 
containing either of $\tchi^0_3, \tchi^0_4, \tchi^\pm_2$.
We thereafter computed with ISASUSY \cite{isa}
the BR of the charginos/neutralinos for the 
considered model and different values of $m(A)$, and convoluted 
the results with the total SUSY cross-section calculated at NLO 
with PROSPINO \cite{PROSPINO}. The expected numbers of events as a function of 
$m(A)$ for an integrated luminosity of 300~fb$^{-1}$ are 
shown in Figure~\ref{fig:plothb}. The curve clearly shows the 
drop in number of events when the decays to $\tchi^0_2$ and
$\tchi^\pm_1$ become kinematically inaccessible, and the 
fast decrease to zero after 300~GeV.
For $m(A)=300$~GeV, the number of 
events contributing to the $b\bar{b}$ peak
is $\sim$1500. It should therefore 
be possible to put a lower bound on the heavy Higgs mass of 
approximately 300 GeV. In order to verify whether this is actually
possible, a detailed experimental analysis is needed, outside
of the scope of the present study.\par

For the SUSY decays to heavy Higgses, two channels have been 
identified as particularly promising, and have been the subject of
detailed experimental analyses 
\cite{atltdr,Bisset:2003ix,Moortgat:2001pp,Hansen:2005fg}:
\begin{itemize}
\item
$H^\pm\rightarrow\tchi^0_i \tchi^\pm_1 \rightarrow 3\ell + E_T^{miss}$
\item
$A/H\rightarrow\tchi^0_2\tchi^0_2\rightarrow 4\ell +  E_T^{miss}$
\end{itemize}
The first channel is hopeless for the considered model, as 
the value of the branching ratio for the charged Higgs decay 
to three leptons through a chargino-neutralino pair  
is  of order a few  10$^{-6}$.\par
For the channel 
$A/H\rightarrow\tchi^0_2\tchi^0_2$, the total BR to 4 leptons through 
the second neutralino is 1.36$\times 10^{-4}$ (0.46$\times 10^{-4}$) respectively for the $A$ ($H$).
We considered the $A$ and $H$ production both through gluon fusion, 
and in association with two $b$ quarks, based on the available NLO
calculations \cite{Spira:1995rr,Dittmaier:2003ej}. 
A total of 40 events per experiment is produced 
for an integrated luminosity of 300~fb$^{-1}$. 
The  typical analysis efficiency for this signal
is 30\%, with the reduction of the SM background to negligible
levels. The dominant SUSY background can be further reduced
with respect to previous analyses by the possibility of
performing a full reconstruction of the A/H mass peak, as discussed in 
\cite{Nojiri:2003tu}. 
It is therefore thinkable  that  the heavy Higgs 
can be discovered in this channel. However,
given the small number of events considered, a very careful background study,
taking into account the details of the detector performance would
be required. \par
In conclusion, three scenarios can be envisaged:
\begin{itemize}
\item
A scenario in which the only constraints on the $m(A)-\tan\beta$ plane
are provided by the measurement of the light Higgs $h$, and by the
non-discovery of the heavy Higgses in their Standard Model decay modes
\item
A scenario in which a lower limit of approximately 300~GeV on the
$H/A$ mass can be set by the non-observation of the heavy Higgses
in the SUSY cascade decays. 
\item
A scenario in which the $H/A$ is discovered in its $\tchi^0_2\tchi^0_2$
decay mode.
\end{itemize}
In the following we will evaluate the constraints on \mbox{$\Omega_\chi h^2$}
taking into account all the three scenarios.

\section{Calculation of the relic density}
We can at this point calculate using the micrOMEGAs program 
\cite{Belanger:2001fz} the LSP relic density \mbox{$\Omega_\chi h^2$}
for each of the Monte Carlo experiments defined in Section \ref{Sec:meas}.
The parameters derived 
from the measurements given in Table~\ref{tab:summes} used to 
create the Monte Carlo 
experiments are the masses of $\tchi^0_1$, $\tchi^0_2$, $\tchi^0_4$,
$\tl_R$, $\tl_L$, and  $\ttau_1$ and the stau mixing angle $\theta_\tau$.
From these, for each experiment the soft MSSM parameters to give
as input to micrOMEGAs are calculated.
The nominal values are assumed for all the other MSSM 
parameters, in particular $m(A)$, $\tan\beta$ and $m(\ttau_2)$. \par
\begin{figure}[htb]
\begin{center}
\dofigs{0.5\textwidth}{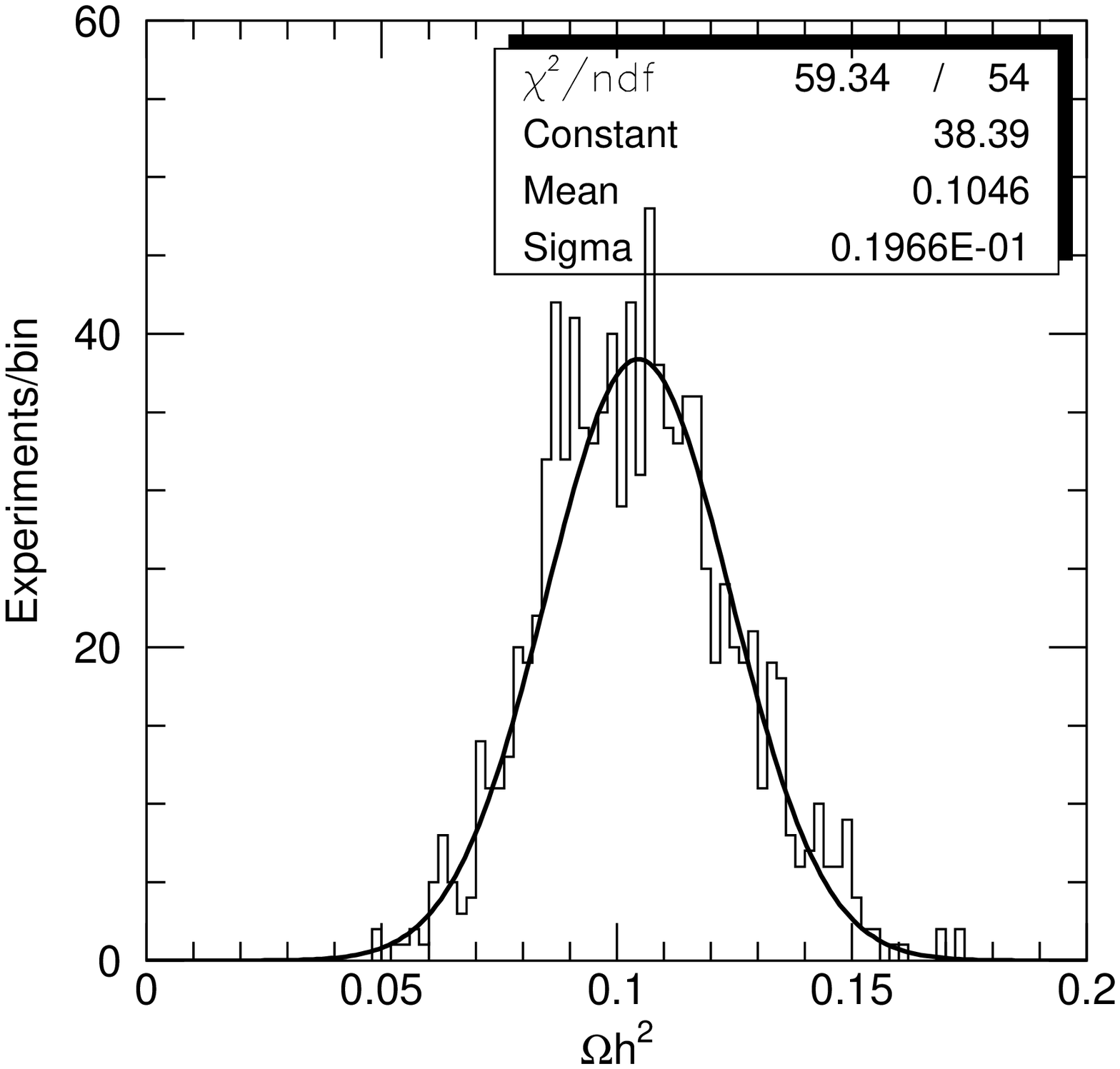}{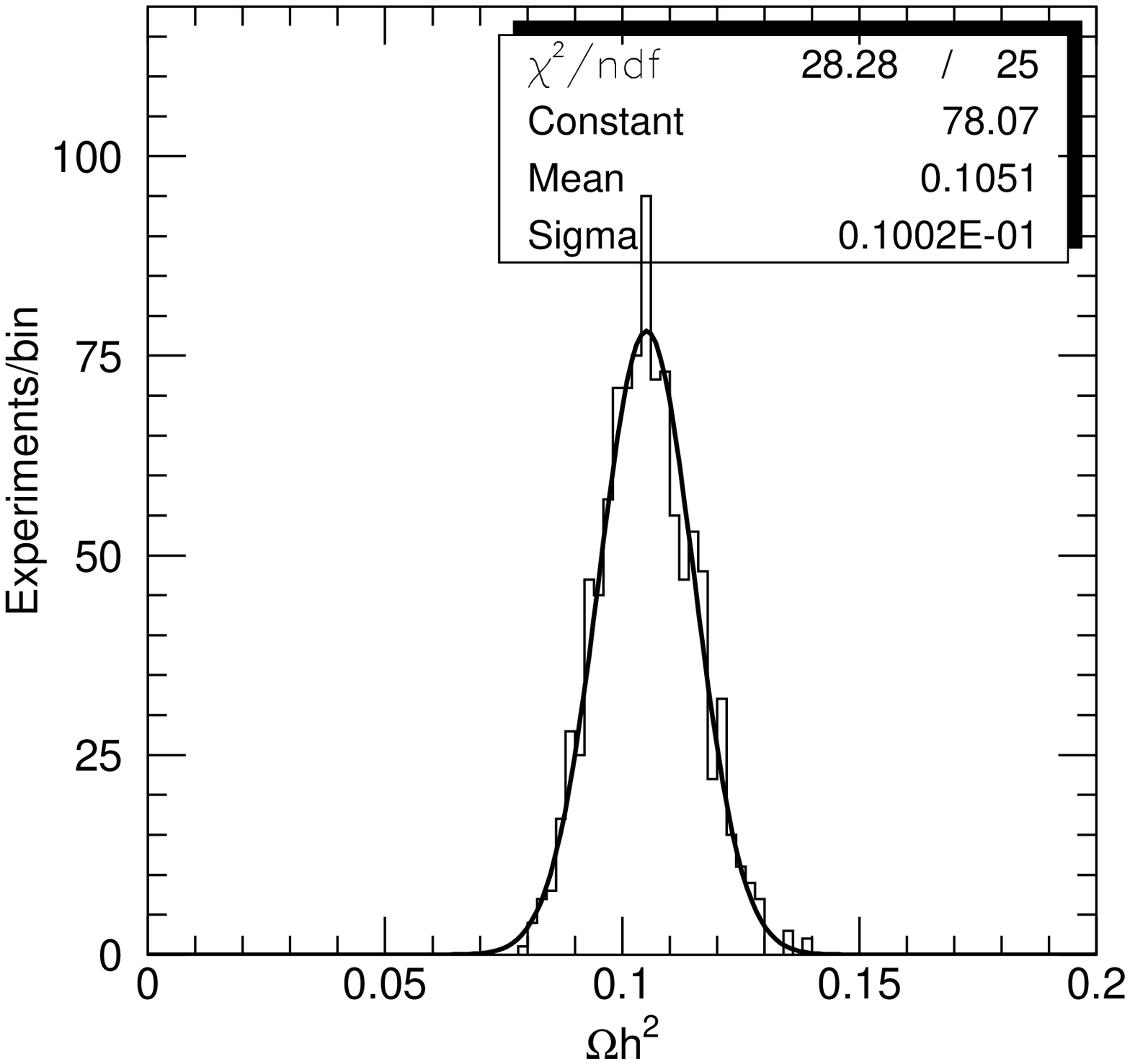}
\caption{\label{fig:omegaexp} {\it 
Distributions of the predicted relic density \mbox{$\Omega_\chi h^2$}
incorporating the experimental errors. The distributions are 
shown for an assumed error on the $\tau\tau$ edge respectively 
of 5 GeV (left) and 0.5 GeV (right).
}}
\end{center}
\end{figure}
\begin{figure}[htb]
\begin{center}
\dofig{0.6\textwidth}{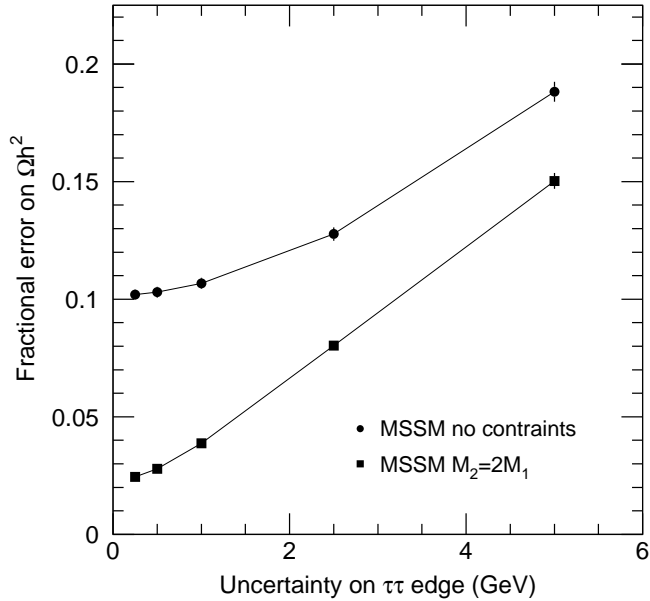}
\caption{\label{fig:omegaerr} {\it 
Fractional uncertainty on the predicted relic density \mbox{$\Omega_\chi h^2$}
from the experimental measurement, as a function of the assumed 
uncertainty on the position of the $\tau\tau$ edge respectively for 
the completely unconstrained MSSM (black circles), and under 
the assumption of gaugino mass unification (black squares). 
}}
\end{center}
\end{figure}

The resulting distributions of the calculated \mbox{$\Omega_\chi h^2$}
are given in Figure~\ref{fig:omegaexp} for two values of the assumed
uncertainty on the position of the $\tau\tau$ edge, 
respectively 5 GeV and 0.5~GeV.
The error is respectively $\sim$20\% (10\%) for an uncertainty  of 5(0.5)~GeV. 
This uncertainty is quoted in Table~\ref{tab:summes} as 5~GeV,  
and translates to an uncertainty on the mass difference
$m(\ttau_1)-m(\tchi^0_1)$ of 2.5 GeV. The measurement precision 
on this  difference has been found 
in \cite{Allanach:2004xn} to be the dominant factor in 
the determination of the relic density.
No detailed study is available on the precision with which 
the LHC experiments can define the $\tau\tau$ edge, and 5~GeV was
quoted in \cite{LHCLC} as a conservative upper limit on this figure. 
It is therefore interesting to treat this 
uncertainty as a parameter of the analysis. We have therefore re-evaluated
the neutralino relic density as a function of the error on  the $\tau\tau$
edge position. The results are shown as
full circles in Figure~\ref{fig:omegaerr}. 
The overall uncertainty depends on the $\tau\tau$ measurement 
as long as the uncertainty is above $\sim1$~GeV, which therefore
should be taken as the goal for the systematic control to be achieved
on the measurement of this variable. \par
\begin{figure}[htb]
\begin{center}
\dofig{0.6\textwidth}{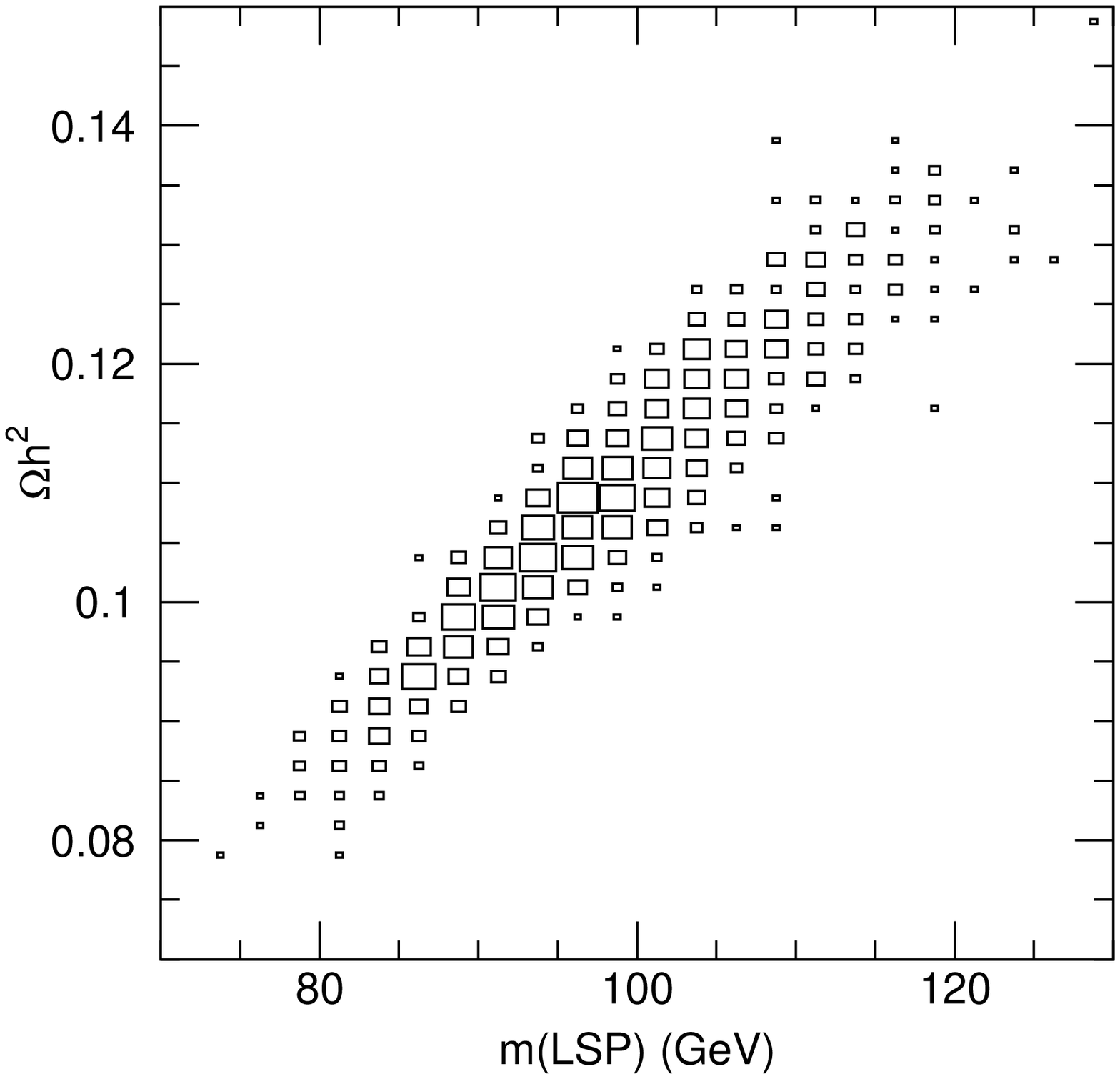}
\caption{\label{fig:chi0dep} {\it
Value of the predicted relic density \mbox{$\Omega_\chi h^2$} as a function
of the measured $\tchi^0_1$ mass. 
}}
\end{center}
\end{figure}

The remaining uncertainty of $10\%$ is determined by the $\sim$10\%
error on the $\lsp$ mass, as can be seen in 
Figure~\ref{fig:chi0dep},
produced for $\sigma(m(\tau\tau))=1$~GeV,
where the predicted relic density is shown as a function of 
the measured $\lsp$ mass.  If we assume gaugino mass universality,
as discussed in Section~\ref{sec:mssm}, the uncertainty on the
$\lsp$ mass is reduced to $\sim250$~MeV. We show as full squares
in  Figure~\ref{fig:omegaerr} the evolution of the uncertainty 
as a function of the error on  the $\tau\tau$
edge position under this assumption. For an uncertainty on $m(\tau\tau)$
of 1~GeV, the measurement error is reduced to $\sim4\%$.\par
We have also studied the effect of changing the systematic uncertainty
on the measurement of $BR(\tilde\chi^0_2\rightarrow\tilde{\ell_R}\ell)/
BR(\tilde\chi^0_2\rightarrow\ttau_1\tau)$ from 10\% to 1\%. 
The effect has been shown to be negligible as compared to the other
uncertainty sources.\par
We further studied  in detail  the dependence of the relic density 
prediction on the parameters which are only very loosely 
constrained (if at all) by the LHC data.  
The squark and gluino masses, will be measured with a precision 
of a few percent at the LHC, but for the masses considered, the
relic density is essentially independent from their value. 
The only loophole could be a light stop. 
We have studied the dependence of the predicted relic density 
on the mass of the stop. We found that a light stop would only contribute to the
relic density if its mass were below  $\sim$140~GeV.
A detailed study on a similar model point, given in 
\cite{Hisano:2003qu}, shows that the lighter stop, which has a mass of $\sim400$~GeV, 
can  be observed in the cascade decays of the gluino, 
and it is possible to build kinematic edges which can be used 
to constrain its mass. We can therefore assume 
that for the considered model it will be possible to exclude a
significant stop coannihilation contribution to the relic density.
The main loosely constrained parameters which influence the relic density
calculation are therefore: $\tan\beta$, $m(A)$, and $m(\ttau_2)$.
We have therefore varied 
in turn $m(\ttau_2)$, $\tan\beta$,  $m(A)$, while the other soft 
parameters going in input to micrOMEGAs are recalculated 
in such a way that the masses of the sparticles which can be
measured experimentally are kept fixed at their nominal value.\par
\begin{figure}[htb]
\begin{center}
\dofigs{0.5\textwidth}{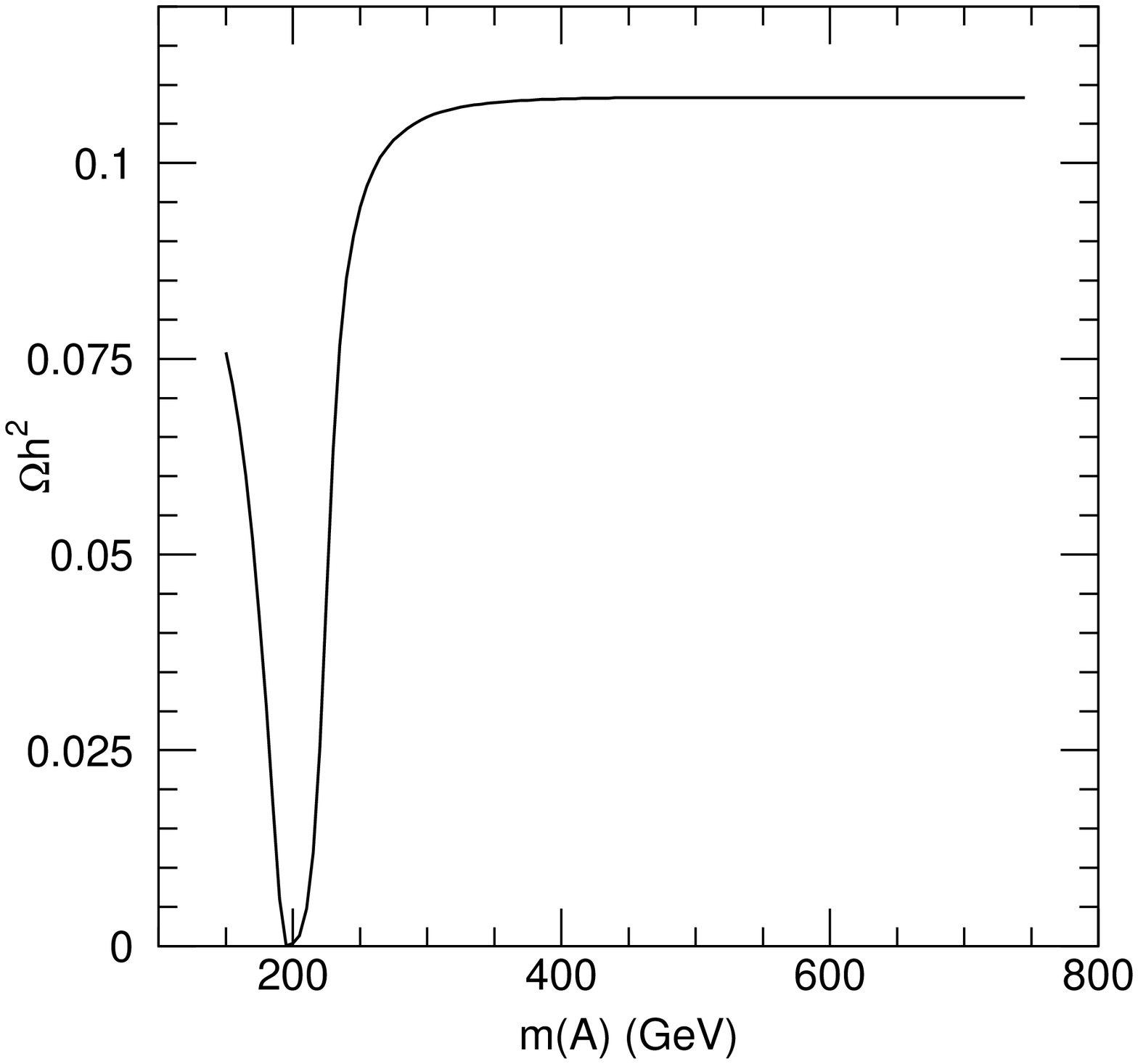}{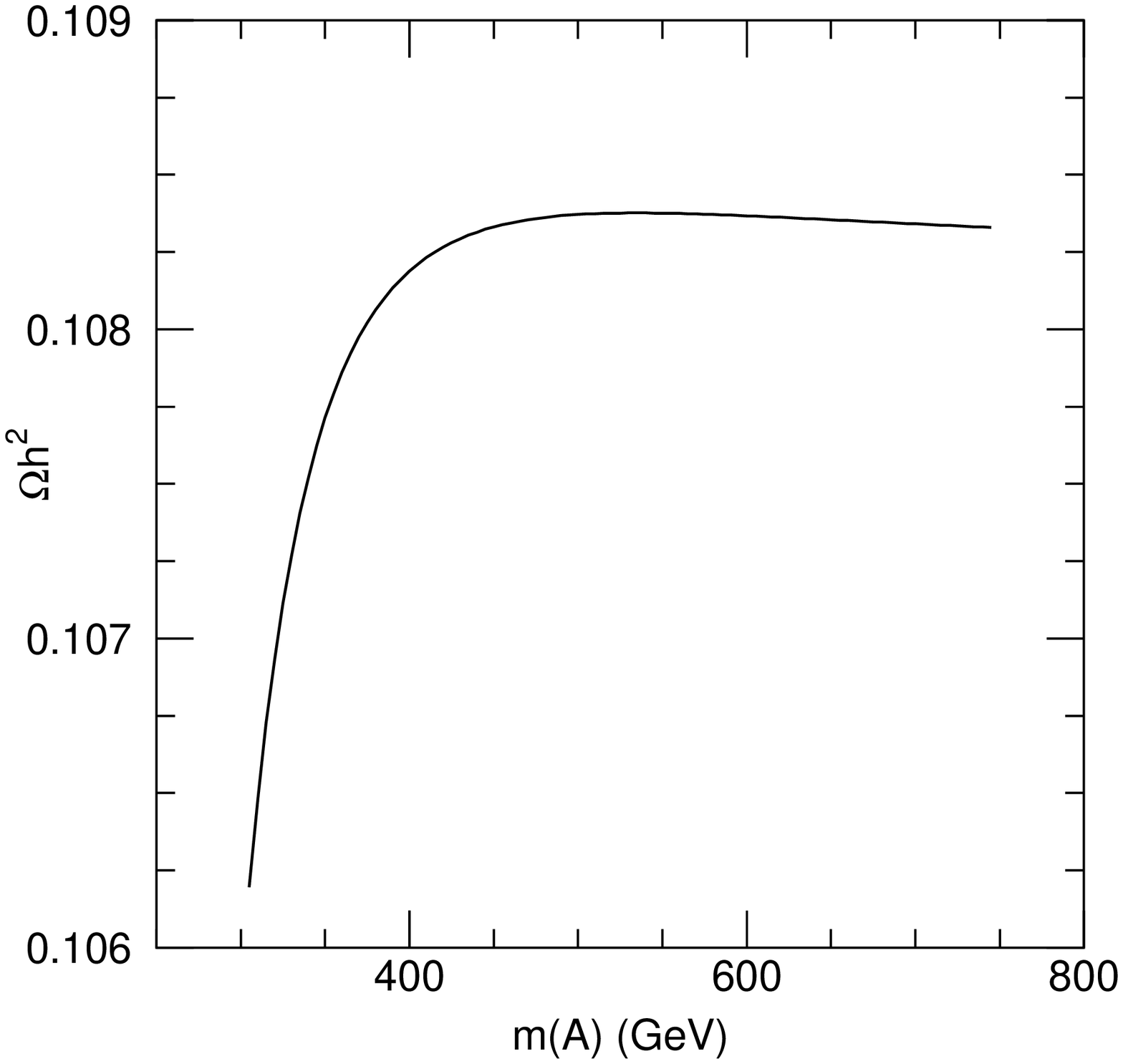}
\caption{\label{fig:omvsma} {\it Dependency of the relic density 
\mbox{$\Omega_\chi h^2$}
from  the value  of the pseudoscalar Higgs mass $m(A)$. The 
dependency is shown respectively for the whole $m(A)$ mass range 
(left), and for $m(A)>300$~GeV (right). 
}}
\end{center}
\end{figure}

\begin{figure}[htb]
\begin{center}
\dofigs{0.5\textwidth}{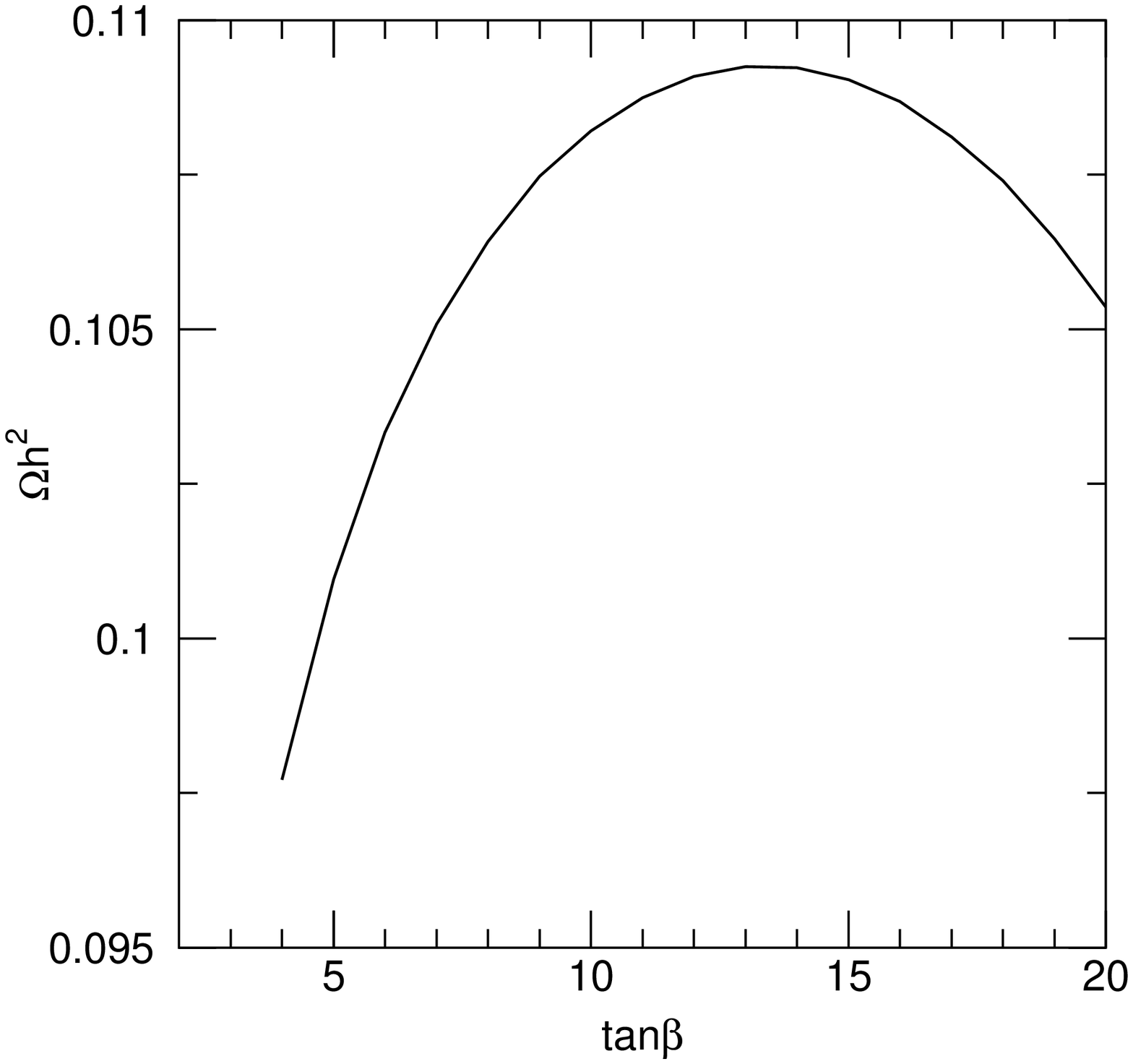}{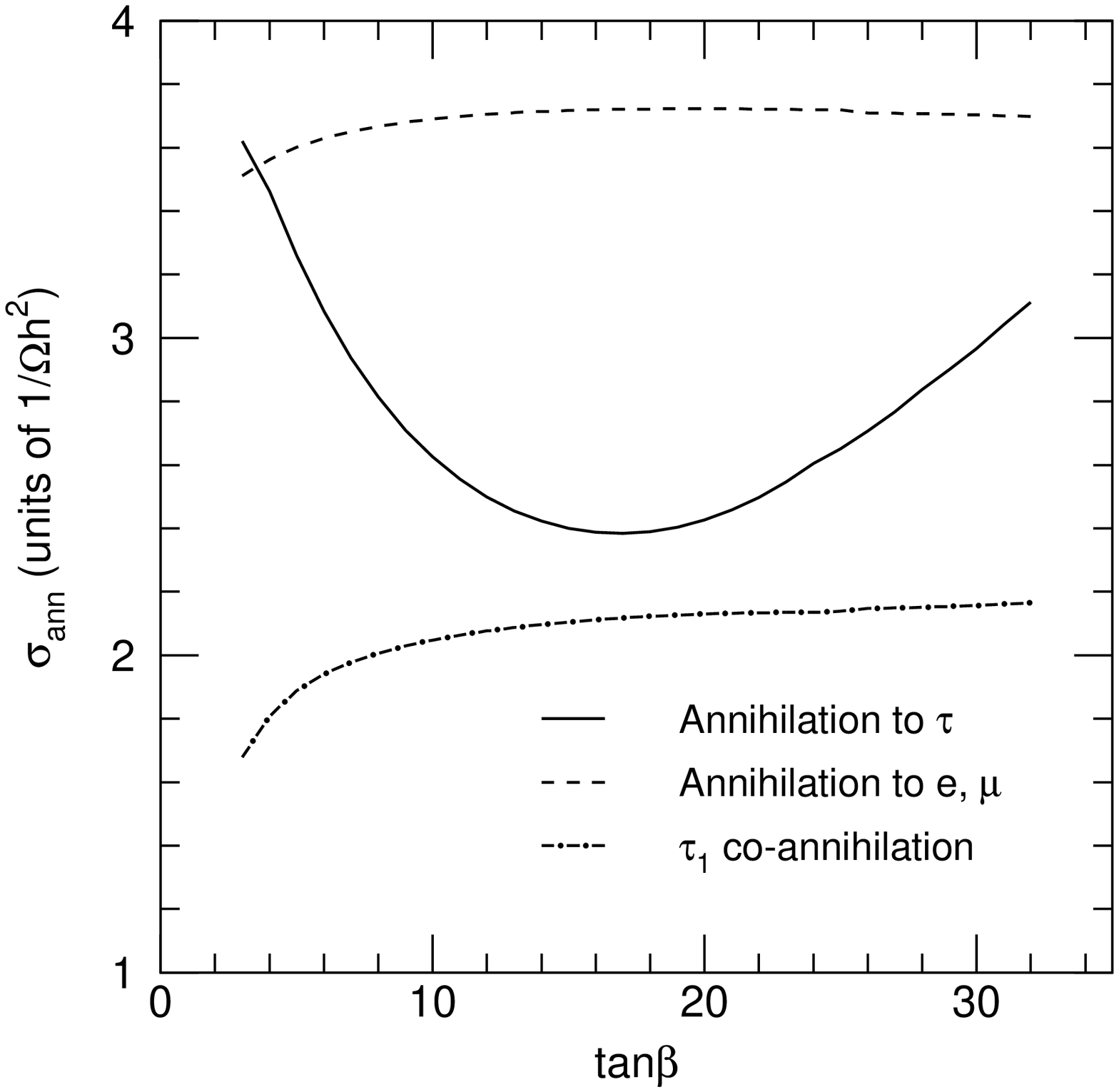}
\caption{\label{fig:omvstb} {\it 
Left:  Dependency of the relic density 
\mbox{$\Omega_\chi h^2$}
from  the value  of $\tan\beta$.
Right: Dependency of the annihilation cross-section for 
different processes, respectively 
$ \lsp \lsp \rightarrow \tau^+ \tau^-$ (full line),
$ \lsp \lsp \rightarrow \ell^+\ell^-$ (dashed line) and
$ \lsp \ttau_1 \rightarrow Z/A \tau$ (dot-dashed line).
The units are contributions to $1/\Omega h^2$.
}}
\end{center}
\end{figure}

\begin{figure}[htb]
\begin{center}
\dofig{0.6\textwidth}{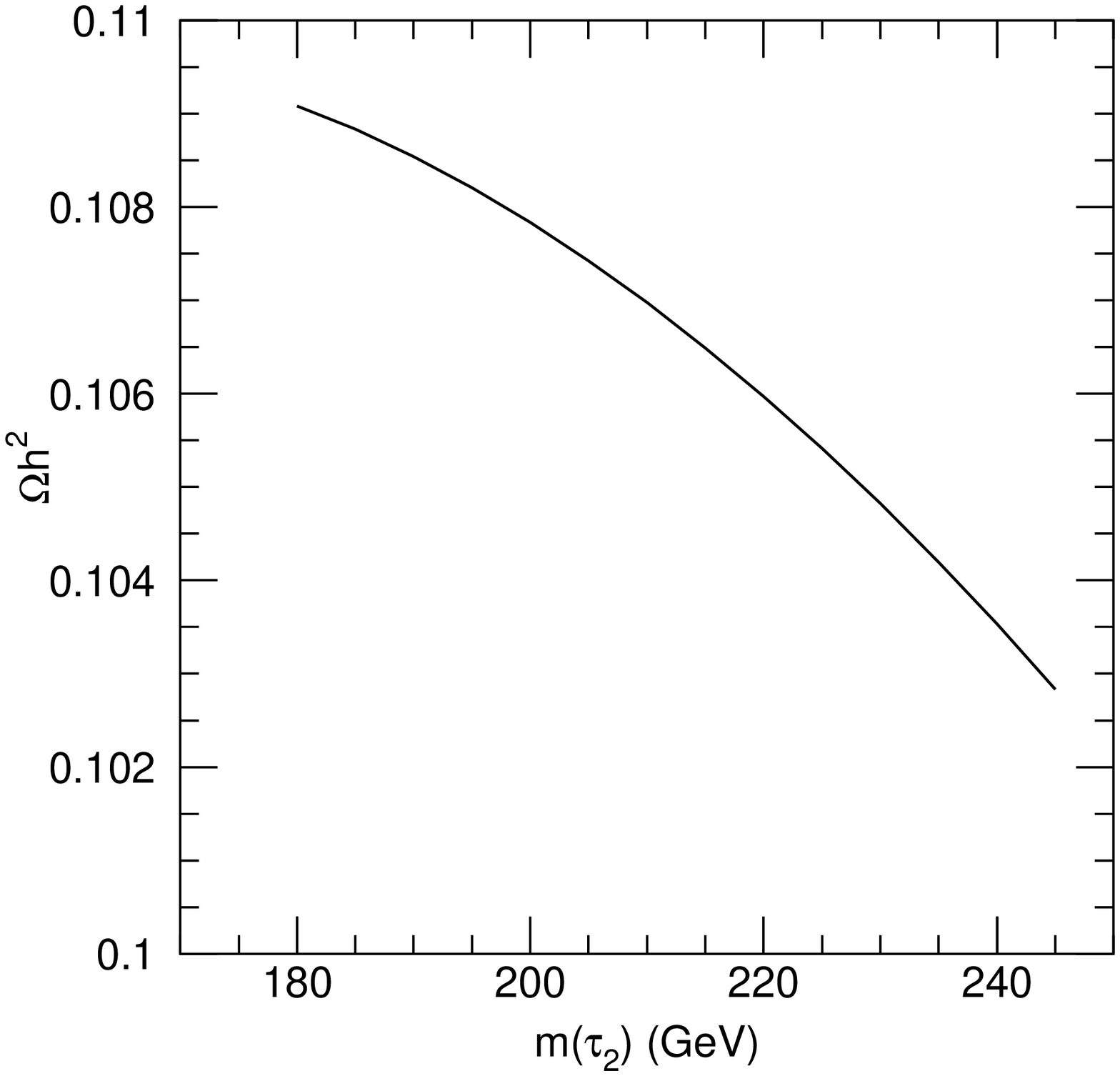}
\caption{\label{fig:omvst2} {\it 
 Dependency of the relic density
\mbox{$\Omega_\chi h^2$}
from  the value  of the $\ttau_2$ mass.
}}
\end{center}
\end{figure}

 
For this exercise we have assumed the  minimum 
constraint in the $m(A)-\tan\beta$ plane discussed above, 
assuming a lower bound on $\tan\beta$ of 3, and  vetoing the regions where
the heavy Higgs can be discovered by the LHC experiments at 5 $\sigma$
through their $SM$ decays. 
We have moreover imposed the loose 
bounds on $m(\ttau_2)$ discussed above.
The dependencies on the different
parameters  are shown in Figures~\ref{fig:omvsma}, \ref{fig:omvstb} and
\ref{fig:omvst2}.

It is clear from these figures that if we can not extract from the
data any direct information on the mass of the heavy Higgses, only 
an upper limit on the neutralino relic density can be given. 
If it is possible to exclude heavy Higgs masses lighter than 300~GeV
through the peak search in the $b\bar{b}$ mass distribution, 
the spread in relic density measurement is of order 1\%, and is
strongly dependent on the experimental value of the lower limit 
on the mass of the heavy Higgs. 
If the heavy Higgses are  directly observable through their decays
to SUSY particles, the Higgs mass is fixed, and its experimental
uncertainty does not contribute to the error on the relic density 
prediction, as for high Higgs masses the contribution of channels
involving Higgses to the neutralino annihilation is negligible.

In this case the dominant contribution to the uncertainty will come
from the poorly constrained value of $\tan\beta$. In the interval
allowed by the non-observation of the $SM$ decays of the Higgs, the
relic density varies by $\sim$11\%, as shown in the right side of
Figure~\ref{fig:omvstb}. It is interesting to disentangle the
contributions of the different annihilation processes to the
variation. In the right side of Figure~\ref{fig:omvstb} we show the
annihilation cross-sections (in units of their contribution to
$1/\Omega$). For the different processes.  The spread is dominated by
the $ \lsp \lsp \rightarrow \tau^+ \tau^-$ process. The reason is that
for each value of $\tan\beta$ we recalculate the soft parameters in
such a way that the sparticle masses, and the branching ratios which
are measured experimentally are kept constant. Therefore the
composition of the $\tchi^0_1$ and the value of $\theta_\tau$ vary as
shown in Figures~\ref{fig:z4dep}, producing the dependency observed in
the full line in the left side of Figure~\ref{fig:omvstb}.\par An
additional uncertainty will come from the value of $m(\ttau_2)$, which
contributes a $\sim7\%$ spread to the result, as shown in
Figure~\ref{fig:omvst2}.  This is because the $\tilde{\tau}_2$
exchange contribution is opposite to the $\tilde{\tau}_1$
contribution.  The cancellation appears in the $s$-wave part of the
pair annihilation cross section, which is chirality suppressed.  In
the limit where the Higgsino component of the lightest neutralino can
be ignored, the $\tilde{\tau}_i$ contribution to the amplitude can be
expressed as
\begin{equation}
M(swave) \propto \sin\theta_{\tau}\cos\theta_{\tau}
[1/(1+m_{\tilde{\tau}_1}^2/m_{\tilde{\chi}^0_1}^2)
- 1/(1+ m_{\tilde{\tau}_2}^2/m_{\tilde{\chi}^0_1}^2)]
Z_{11}^2
\end{equation}
yielding the observed dependence of the annihilation cross-section
on $m(\ttau_2)$.
The mixing angle $\theta_{\tau}$ is kept constant by construction
in  Figure~\ref{fig:omvst2}  because we use the ratio of decay widths
\mbox{
$\Gamma(\tilde{\chi}^0_2\rightarrow ee\tilde{\chi}^0_1)/
\Gamma(\tilde{\chi}^0_2\rightarrow \tau\tau\tilde{\chi}^0_1)$}
as the constraint on $\theta_\tau$.

\section{Discussion}
In the above discussion we have considered a `bulk region' SUSY model
in which $\lsp$ annihilation in the early universe is dominated by
diagrams involving light sleptons. This model lies within the mSUGRA
sub-space of the full MSSM parameter space, however we have shown that
it is possible to set useful constraints on the neutralino relic
density without making special assumptions on the behaviour of the
theory at high scale.

It is interesting to consider whether this result applies more
generally to SUSY models characterised by dominant light slepton
$\lsp$ annihilation. The starting point in the analysis is the
possibility of isolating the decay chains $\tq_L\to\tchi^0_2
q\to\tl_r^\pm\ell^\mp q\rightarrow\ell^\pm\ell^\mp q$
$\tq_L\to\tchi^0_2 q\to\ttau_1^\pm\tau^\mp
q\rightarrow\tau^\pm\tau^\mp q$ with sufficient statistics to be able
to observe the kinematic edges providing the mass measurement.  One
therefore requires the mass hierarchy $m(\tq_L)> m(\tchi^0_2)
>m(\tl_R)$ and equivalently for the $\ttau_1$. The former condition on
$m(\tchi^0_2)$ and $m(\tl_R)$ is generally satisfied in the light
slepton annihilation region, and if stau annihilation is to be
relevant so must the latter. To avoid significant squark
co-annihilation $m(\tq_L)$ must by definition be somewhat larger than
the mass scale of the sleptons and lightest neutralino. It is
therefore likely that at least the necessary decay chain(s) will
occur, although the mass differences may be sufficiently small that
one or more of the kinematic end-points are unobservable.

In order to obtain sufficient statistics of events containing the
above decay chain, $m(\tq_L)$ must be $\lesssim$ 1 TeV, and the
$\tchi^0_2$ must have a large Wino component to couple to
$\tq_L$. Moreover, the mass of the left handed component of $\ttau_1$
should not be too large, to avoid the decay
$\tchi^0_2\rightarrow\ttau_1\tau$ saturating the $\tchi^0_2$ branching
ratio and thus killing the lepton signature. From the experimental
point of view, we need to add the requirement that the masses of the
involved sleptons are not too near to the mass of either neutralino.
A further essential ingredient in the reconstruction of the neutralino
mass matrix is the possibility of measuring the mass of the
$\tchi^0_4$. This implies non-negligible gaugino components in the
$\tchi^0_4$, the appropriate mass hierarchy and typically the
knowledge of the $\tl_L$ mass, either through its direct production or
through it appearance in a cascade decay \cite{Polesello:2004aq}. Such
requirements on the neutralino and squark sectors need not be
satisfied in the light slepton $\lsp$ annihilation region; if this
were indeed the case the analysis would be more challenging.

Finally, we also require that it be possible to constrain alternative
contributions to the annihilation cross-section through exclusion of
the required values of $m_A$ and $m(\ttop_1)$ by direct measurement or
non-observation of a signal in an appropriate channel. If such
constraints can not be obtained then the obtained value for $\Omega
h^2$ is merely an upper limit.

We may consider also whether the techniques described here may be
applied to SUSY models in which alternative annihilation mechanisms
dominate. In general in order to obtain an estimate of $\Omega h^2$ it
is necessary not just to measure the dominant contributions to the
$\lsp$ annihilation cross-section but also to constrain all other
possible contributions. In any such analysis therefore the masses and
mixings of the neutralino, light slepton, stau, stop and Higgs
sectors must all be measured or constrained. The goals of the analysis
must therefore be similar to that described here and if accessible use
will be made of the decay chains considered above.

Although it is impossible to outline the required analyses for all
possible models a few general observations may be made. Models in
which stau co-annihilation dominates display similar phenomenology to
the light slepton annihilation region, however the small mass
difference between the $\ttau_1$ and $\lsp$ makes observation and
measurement of the $\tq_L\to\tchi^0_2 q\to\ttau_1^\pm\tau^\mp
q\rightarrow\tau^\pm\tau^\mp q$ decay chain more difficult. In regions
in which the $\lsp$ possesses a significant Higgsino or Wino component
(see e.g. \cite{Baer:2005zc,Baer:2005jq}), leading to dominant
(co-)annihilation to EW bosons the analysis would be qualitatively
different to that described here and highly model dependent. If heavy
Higgs annihilation is important techniques such as those described
above for measuring its mass will be vital. When stop co-annihilation
is enhanced the small mass difference between the strongly interacting
$\ttop_1$ and $\lsp$ will likely make measurement of the masses very
difficult indeed.

\section{Conclusions}

We have explored the potential of the LHC experiments for predicting
the cosmological relic density of the LSP from detailed measurements
of the SUSY spectrum and decay modes. No unification condition is
imposed on the sparticle spectrum.  We have focused on a model with
essentially bino LSP and light sleptons, for which a detailed
experimental study exists in the literature.  We have examined the
relative roles of the different measurement uncertainties, and studied
the uncertainties due to SUSY parameters which are relevant for the
relic density calculation, and are poorly, if at all, constrained by
the measurements at the LHC.  For the experimentally accessible
measurements, we have highlighted the role of the measurement of the
$\tau\tau$ edge. For the influence of badly measured parameters, the
key issue is the observability of the heavy Higgses at the LHC. We
have discussed in particular the prospect for observing the H/A in the
cascade decays of the sparticles or in a SUSY decay mode.  In case the
position of the $\tau\tau$ edge can be controlled at the level of
$\sim$1~GeV, and the LHC experiments can demonstrate that the Heavy
Higgses $H,A$ have a mass in excess of 300~GeV, the value of the
neutralino relic density can be predicted as:
$$ \Omega_\chi h^2 = 0.108 \pm 0.01 (stat + sys)^{+0.00}_{-0.002}
(M(A)) ^{+0.001}_{-0.011} (\tan\beta) ^{+0.002}_{-0.005} (m(\ttau_2))
$$ 
after three years of data taking at high luminosity, corresponding
to an integrated luminosity of 300~fb$^{-1}$.  In case no experimental
information on the heavy Higgs can be extracted from the LHC data, it
will only be possible to put an upper limit of approximately 0.12 on
the neutralino relic density.  The
discovery of the decay of the heavy Higgses to SUSY particles is 
however most probably statistics limited, and a luminosity upgrade 
of the LHC might allow the discovery of this decay mode, thus 
yielding also a constraint on $\tan\beta$.

\section*{Acknowledgments}
We would like to thank Genevieve Belanger for useful discussions on 
the usage of micrOMEGAs.

\end{document}